\documentclass[12pt]{book}

\usepackage{a4wide}
\usepackage{fancyhdr}
\usepackage{amssymb,amsmath}
\usepackage{bm}
\DeclareMathOperator{\Tr}{Tr}
\usepackage[skip,indent]{parskip}
\usepackage{graphicx}
\usepackage{rotating}
\usepackage[dvipsnames,table,xcdraw]{xcolor}
\usepackage{subcaption}
\usepackage[nottoc,numbib]{tocbibind}

\usepackage[bookmarks]{hyperref}

\newcommand{\D}[2][\phantom{}]{\ensuremath{\text{D}#2_{#1}}}
\newcommand{\F}[1][\phantom{}]{\ensuremath{\text{F1}_{#1}}}
\newcommand{\W}[1][\phantom{}]{\ensuremath{\text{P}_{#1}}}
\newcommand{\NS}[1][\phantom{}]{\ensuremath{\text{NS5}_{#1}}}
\newcommand{\KKM}[1][\phantom{}]{\ensuremath{\text{KKM}_{#1}}}
\newcommand{\M}[2][\phantom{}]{\ensuremath{\text{M}#2_{#1}}}

\usepackage{graphicx}
\usepackage{tikz}
\usetikzlibrary{decorations.markings}
\usetikzlibrary{shapes,shapes.geometric, arrows, positioning}
\usetikzlibrary{fadings}
\usetikzlibrary{external}
\usetikzlibrary{tikzmark}
\tikzset{->-/.style={decoration={
			markings,
			mark=at position #1 with {\arrow{stealth}}},postaction={decorate}}}
\tikzset{%
            base/.style = {rectangle, rounded corners, draw=black,
                           minimum width=2cm, minimum height=.8cm,
                           text centered},
             IIA/.style = {base, fill=red!20},
             IIB/.style = {base, fill=yellow!20},
            IIAB/.style = {base, fill=orange!20},
        TDuality/.style = {thick,->,>=stealth},
        SDuality/.style = {gray!70,thin,<->,>=stealth},
        CommutingLine/.style = {cyan},
        AnticommutingLine/.style = {orange,dashed,thick},
}
\usepackage{adjustbox}
\usepackage{pgfplots}
\pgfplotsset{compat=1.11}
\usepgfplotslibrary{fillbetween}
\usetikzlibrary{intersections}
\usetikzlibrary{patterns}

\pgfdeclarelayer{bg}
\pgfsetlayers{bg,main}

\renewcommand{\chaptermark}[1]%
         {\markboth{\thechapter.\ #1}{}}
\renewcommand{\sectionmark}[1]%
         {\markright{\thesection\ #1}}
\newcommand{\LMUTitle}[9]{
  \thispagestyle{empty}
  \vspace*{\stretch{1}}
  {\parindent0cm
   \rule{\linewidth}{.7ex}}
  \begin{flushright}

    \vspace*{\stretch{1}}
    \sffamily\bfseries\Huge
    #1\\
    \vspace*{\stretch{1}}
    \sffamily\bfseries\Large
    (#2)\\
    \vspace*{\stretch{1}}
    \sffamily\bfseries\large
    #3
    \vspace*{\stretch{1}}
  \end{flushright}
  \rule{\linewidth}{.7ex}
  \vspace*{\stretch{5}}
  \begin{center}
    \includegraphics[width=2in]{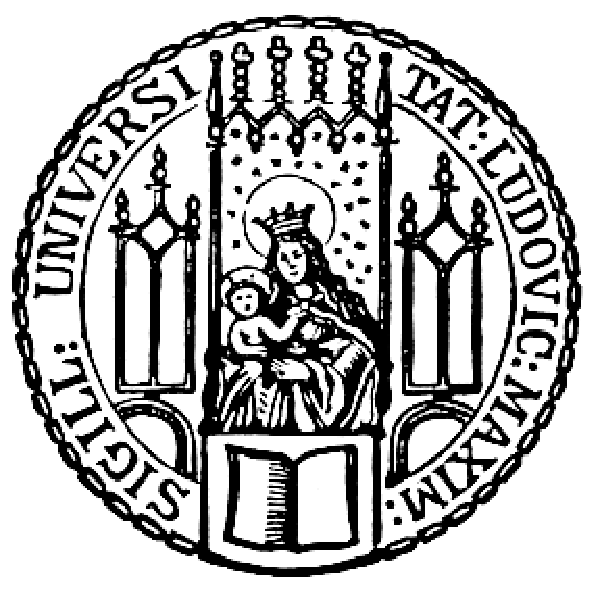}
  \end{center}
  \vspace*{\stretch{1}}
  \begin{center}\sffamily\LARGE{#6}\end{center}
  \newpage
  \thispagestyle{empty}

  \cleardoublepage
  \thispagestyle{empty}

  \vspace*{\stretch{1}}
  {\parindent0cm
  \rule{\linewidth}{.7ex}}
  \begin{flushright}
    \vspace*{\stretch{1}}
    \sffamily\bfseries\Huge
    #1\\
    \vspace*{\stretch{0.5}}
    \sffamily\bfseries\Large
    (#2)\\
    \vspace*{\stretch{1}}
    \sffamily\bfseries\large
    #3
    \vspace*{\stretch{1}}
  \end{flushright}
  \rule{\linewidth}{.7ex}

  \vspace*{\stretch{3}}
  \begin{center}
    \Large Master Thesis\\
    \Large at the #5\\
    \Large at the Ludwig Maximilian University\\
    \Large of Munich\\
    \vspace*{\stretch{1}}
    \Large submitted by\\
    \Large #3\\
    \Large from #4\\
    \vspace*{\stretch{2}}
    \Large Munich, #7
  \end{center}

  \newpage
  \thispagestyle{empty}

  \vspace*{\stretch{1}}

  \begin{flushleft}
    \large First Supervisor:  #8 \\[1mm]
    \large Second Supervisor: #9 \\[1mm]
  \end{flushleft}

  \cleardoublepage
}

\newcommand{\Declaration}[2]{
    \chapter*{Declaration of Authorship}
    Declaration:
    I hereby declare that this thesis is my own work, and that I have not used
    any sources and aids other than those stated in the thesis.
    
    \vspace{1cm}
    
    \renewcommand{\arraystretch}{0} 
    
    \begin{flushright}
    	\begin{tabular}{rr}
    		München, #2, & \hspace*{5cm}\\[0mm]
    		\cline{2-2}\\[2mm]    
    		& #1       
    	\end{tabular}
    \end{flushright}
    
    \renewcommand{\arraystretch}{1}
    
    \cleardoublepage
}

\begin{document}

  \frontmatter

  \LMUTitle
      {Local Supersymmetry Enhancement\\and Black Hole Microstates}                                                          
      {Erweiterung der lokalen Supersymmetrie\\und Mikrozustände von schwarzen Löchern}                                 
      {Ben Eckardt}                            
      {Aalen}                                  
      {Faculty of Physics}                     
      {Munich 2024}                            
      {14.05.2024}                             
      {PD Dr. Ralph Blumenhagen}               
      {Dr. Yixuan Li}                          

  \chapter*{Abstract}
\addcontentsline{toc}{chapter}{\protect Abstract}

In string theory, black holes can be made from brane systems that are partially delocalized in string theory's extra dimensions. It is expected that they correspond to an ensemble of pure, horizonless microstates sourced by localized versions of this brane system. This thesis aims at studying the relationship between supersymmetric black holes and their microstates. In particular, we apply a concept called local supersymmetry enhancement to 1/4-BPS and 1/8-BPS systems that correspond to two-, three-, and four-charge black holes. This concept locally increases the number of preserved supercharges to 16 by adding dipole charges to the system. These bound states are then expected to be horizonless and describe the microstates of the black hole. 
First, we list all dipole charges for 1/4-BPS systems in Type II string theory. Then, we apply these to generate two structures to enhance 1/8-BPS systems, one with exclusively internal dipole charges and the other with a non-trivial shape in the non-compact dimensions. 
Thus, we identify the dipole charges for the microstates of various supersymmetric black holes, including new internal dipole charges for the $\D{1}$-$\D{5}$-$\W$ black hole and external ones for the $\D{2}$-$\D{2}$-$\D{2}$-$\D{6}$ black hole. 
These dipole charges reveal the local structure of the black hole's microstates and will be used in the construction of the corresponding supergravity solutions.

  \Declaration
      {Ben Eckardt}                            
      {14.05.2024}                            

  \tableofcontents

  \listoffigures
  \cleardoublepage

  \mainmatter\setcounter{page}{1}
  \chapter{Introduction}

There are currently two different fundamental theories that describe the real world, general relativity and quantum field theories. The former is a theory of gravity, that describes the cosmos on its larger scales; the latter explains the (usually) short-ranged interactions between fundamental particles on the smallest accessible scales. These two theories are inherently different in that they cannot be unified in a simple way consistently. This poses a problem for systems with a high mass, and density. There, gravity becomes a driving force, and high densities reduce the distance between particles such that short-interaction between the particles become relevant. An extreme example is the black hole. Its gravitational pull is strong enough such that nothing can escape from inside its event horizon, while its mass is concentrated on the smallest possible volume.

In 1974, Stephen Hawking showed by including quantum mechanical effects that black holes emit particles from the horizon. It slowly loses energy by emitting the so-called Hawking radiation at a temperature
\begin{equation}
    T = \frac{1}{8 \pi \, G_N \, M} \,
\end{equation}
where $M$ is the mass of the black hole \cite{Hawking:1974rv}. This leads to the famous information paradox. Suppose we throw an object into the black hole and wait until the black hole radiates away energy equivalent to the mass of the object. As Hawking-radiation is thermal, it does not contain any information on the object. All previous information on the object is lost.

Another puzzle concerns microstates of a black hole. In statistical physics, the microstate of a thermodynamic system, e.g. a fluid or gas, is described by the positions and momenta of the individual particles. As the particles move and interact with each other, these change over time, leading to various microstates for the same system. 
The entropy then measures how many microstates a given system can assume. 
Succeeding the temperature formula, Hawking also proved an equation for the entropy of a black hole in 1975, that was previously conjectured by Jacob Bekenstein \cite{Bekenstein:1972tm,Hawking:1975vcx}. The so-called Bekenstein-Hawking entropy is directly proportional to the area $A$ of the event horizon
\begin{equation}\label{eq:Bekenstein-Hawking-entropy}
    S_\text{BH} = \frac{A}{4 \, G_N} \, .
\end{equation}
The question arises, what are the microstates of the black hole that the Bekenstein-Hawking entropy is counting? 
In general relativity, the black hole is not a thermodynamic system and there are no microstates. This is the result of the no-hair theorem. It states that a black hole is characterized only by its mass, charge and angular momentum \cite{Israel:1967,Israel:1968,Carter:1971}, leaving no degrees of freedom for the microstates. 
Both puzzles, the information paradox and the missing microstates, have to be resolved by a theory of quantum gravity, the unification of general relativity and quantum mechanics.

One contender is string theory. Its fundamental objects are vibrating, open or closed strings. Different particles are described by different excitations of these strings. Among the massless excitations is the graviton, the force carrier of gravity. If the energy of the system is low enough ($E \ll 1/l_S$\footnote{$l_S$ is the length of a string. Its inverse sets the energy scale for strings.}), we can neglect massive excitations of the string. 
The strength of interactions between closed strings is governed by the string coupling $g_S$. In quantum field theories, the cross-section of e.g. electron scattering can be calculated by an expansion in the coupling strength $g$, called perturbation theory, if the coupling strength is weak enough, $g \ll 1$. A similar expansion can be performed in string theory. Quantum effects become relevant if the string coupling $g_S$ is of the order of unity, $g_S \sim 1$, and closed strings start to back-react. 

The open sector of string theory describes strings stretched between so-called D$p$-branes, where $p$ denotes the dimension of the object. The dimension $p$ can be omitted, if it is not relevant. For $N$ D-branes stacked on top of each other, the coupling strength to the strings is given by $g_S N$. 
To create a non-trivial geometries like the black hole, the branes have to back-react on the closed strings, i.e. $g_S N \gg 1$. 
Quantum effects, however, can alter or destroy these geometries. 
To stay within the classical regime $g_S \ll 1$, while creating non-trivial backgrounds, we are considering large stacks of D-branes, $N \gg 1$.

In total, string theory has ten spacetime dimensions. If it describes the real world, six of these dimensions have to be compactified. Black holes in string theory are stacks of D-branes that extend along those compactified directions, e.g. a torus $T^6$. From the point of view of an outside observer in the transverse space, these are then point-like objects curving spacetime. One example of such a configuration is the $\D{2}$-$\D{2}$-$\D{2}$-$\D{6}$ black hole. 
It consists of three orthogonal stacks of $\D{2}$-branes, each spanning two of the six dimensions of the torus $T^6$ while being smeared along the other four directions. Additionally, they are encompassed by a stack of $\D{6}$-branes that fill the whole $T^6$. 
Their total charges $Q_i$ are directly proportional to the number of branes $N_i$ in the stack. 
Solving for the spacetime metric then results in\footnote{Deriving the metric for a four-charge black hole is highly non-trivial. The simple structure in this case follows from a special preserved symmetry called supersymmetry (see below). For more information, see \cite{Maeda:2011sh,Tseytlin:1996bh}.}
\begin{equation}\label{eq:D2D2D2D6-supergravityMetric}
\begin{split}
    ds^2 &= - \left( Z_1 Z_2 Z_3 Z_4 \right)^{-1/2} \, dt^2 + \left( Z_1 Z_2 Z_3 Z_4 \right)^{1/2} \, dr^2 + r^2 \, d\Omega^2 \\
    &+ \left( \frac{Z_2 Z_3}{Z_1 Z_4} \right)^{1/2} \, \left( dx_1^2 + dx_2^2 \right) + \left( \frac{Z_1 Z_3}{Z_2 Z_4} \right)^{1/2} \, \left( dx_3^2 + dx_4^2 \right) + \left( \frac{Z_1 Z_2}{Z_3 Z_4} \right)^{1/2} \, \left( dx_5^2 + dx_6^2 \right) \, ,
\end{split}
\end{equation}
with the warp factors $Z_i = 1 + Q_i / r$ \cite{Tseytlin:1996bh}. Far away from the configuration, $r \to \infty$, this reduces to Minkowski spacetime. For $r \to 0$, the metric has a coordinate singularity, the event horizon. 
With the redefinition $r = \rho - Q_i$, the warp factors become
\begin{equation}
    Z_i = 1 + \frac{Q_i}{\rho - Q_i} = \left( 1 - \frac{Q_i}{\rho} \right)^{-1} \, .
\end{equation}
Together with the assumption of equal charges $Q_i = Q$, this simplifies exactly to the extremal Reissner-Nordström metric. the metric of a charged black hole with mass $M=Q$ in general relativity,
\begin{equation}
    ds^2 = - \left( 1 - \frac{Q}{\rho} \right)^2 \, dt^2 + \left( 1 - \frac{Q}{\rho} \right)^{-2} \, d\rho^2 + \rho^2 \, d\Omega^2 \, .
\end{equation}
In general relativity, this metric describes a charged black hole with mass $M=Q$. 

The solution \eqref{eq:D2D2D2D6-supergravityMetric} is only valid for small curvatures. In general, large curvatures relate to high energies, such that massive excitations of the string become relevant. This low-energy regime of string theory is called supergravity. As the curvature at the event horizon is inversely proportional to $g_S N$, equation \eqref{eq:D2D2D2D6-supergravityMetric} breaks down at $r=0$ for $g_S N \gg 1$. 
Without a proper understanding of non-perturbative string theory and massive excitations of strings, we cannot probe the interior of the black hole directly, but it is possible to calculate its entropy from a microscopic point of view. The key observation is that the entropy does not depend on the string coupling $g_S$. Therefore, we can count the number of microstates of the system in the regime $g_S N \ll 1$ by reducing the string coupling appropriately. This is then no longer a black hole. In fact, the correspondence principle states that when decreasing the string coupling $g_S$, the black hole eventually becomes a state of strings and D-branes with the same charges \cite{Horowitz:1996nw}. 
The resulting system can be described by open strings stretched between stacks of D-branes. The closed string theory sector, and therefore also gravity, decouples. With open string perturbation, the entropy results in
\begin{equation}\label{eq:entropy-D2-D2-D2-D6}
    S \propto \sqrt{N_{\D{2}} N_{\D{2}} N_{\D{2}} N_{\D{6}}} \, .
\end{equation}
This can be compared with the Bekenstein-Hawking entropy in equation \eqref{eq:Bekenstein-Hawking-entropy} by computing the area of the event horizon for the $\D{2}$-$\D{2}$-$\D{2}$-$\D{6}$ metric in equation \eqref{eq:D2D2D2D6-supergravityMetric}, and they do match. This is the first account of a derivation of the black hole entropy by counting microstates. It was first done by Strominger and Vafa for a five-dimensional $\D{1}$-$\D{5}$-$\W$ black hole in \cite{Strominger:1996sh}, and soon generalised to four-dimensional black holes like the $\D{2}$-$\D{2}$-$\D{2}$-$\D{6}$ black hole above in \cite{Maldacena:1996gb,Johnson:1996ga}.

However, the question remains what the black hole microstates look like in the regime of parameters where the black hole exists. 
The main idea this thesis follows is to think of the delocalized brane system as a thermodynamic ensemble. The microstates should then differ slightly from this delocalized brane system, without changing the total charges or the amount of preserved supersymmetry. One idea to implement this is by adding dipole charges to the system, whose net charge vanishes. 
Moreover, we have to make sure that those microstates are pure. From the Bekenstein-Hawking entropy follows immediately that the existence of a horizon implies a mixed state. Thus, the microstates have to be horizonless. Some authors, like in \cite{Bena:2022wpl}, expect systems with 16 local supercharges to be horizonless. We do not have a direct connection between local supercharges and horizons; however, many instances, like the supertube \cite{Mateos:2001qs,Emparan:2001ux} the magnetube \cite{Bena:2013ora,Mathur:2013nja} or superstrata \cite{Bena:2011uw,Bena:2015bea}, support this claim.

To understand the concept of local supersymmetry, we first turn to global supersymmetry. The vacuum in Type II string theories is invariant under 32 different supersymmetry transformations. Introducing strings and branes usually breaks half of them for each type of brane. 
Therefore, the $\D{1}$-$\D{5}$-$\W$ brane system preserves four supersymmetries globally. Those systems are called 1/8-BPS, as they preserve 1/8th of the total supersymmetry. If the system varies in space, the supersymmetries preserved globally are the ones that are preserved everywhere. When zooming in on one specific point, though, the system can seem as if it preserves more supersymmetries. 

Let us illustrate this for an $\F$-$\NS$-$\W$ configuration. 
The entropy of the weakly coupled system ($g_S N \ll 1$) comes from a fractionation of the $N_{\F}$ fundamental strings into $N_{\F} N_{\NS}$ \cite{Seiberg:1997zk,Kutasov:2001uf}. 
This fractionation is best illustrated by going to M-theory, the 11-dimensional superstring theory that reduces to Type IIA string theory when compactified on a small circle. The fundamental strings become $\M{2}$-branes that intersect the $N_{\M{5}}$ $\M{5}$-branes orthogonally. This fractionates the $\M{2}$-branes into $N_{\M{2}} \cdot N_{\M{5}}$ 'little' strips. The entropy is then generated by transverse oscillations of the little strips carrying momentum along the common direction of the $\M{2}$- and $\M{5}$-brane. This translates into $N_{\F} \cdot N_{\NS}$ fractionated 'little' strings oscillating in the four transverse directions of the $\NS$-brane \cite{Dijkgraaf:1996cv}.

When increasing the string coupling $g_S$ and thus the interaction between the strings and D-branes, the fundamental strings start to pull on the $\NS$-branes. In the M-theory uplift, $\M{2}$-branes start to pull on the $\M{5}$-branes, forming a furrow similar to a Callan-Maldacena spike \cite{Callan:1997kz}. This is effectively described by introducing $\M{2}$- and $\M{5}$ dipole charges such that the furrow is still locally 1/2-BPS.  
Furthermore, glues for the $\M{5}$-brane and momentum $\W$ describe the oscillating motion of the $\M{5}$-brane along the M-theory direction, and analogously oscillations of the $\M{2}$-branes along the other directions of the $\M{5}$. The result after the fractionation described above is a maze-like structure that gives a clear idea of the back-reaction of the branes and, thus, also the microstates of the black hole. This back-reacted system is called the super-maze \cite{Bena:2022wpl}.

To summarize, supergravity breaks down in the centre of black holes. Tuning the string coupling such that $g_S N \ll 1$, we can use open string perturbation theory to describe the brane system and investigate the microstates in the weak coupling limit. By including dipole charges, that we call glues, we can investigate the structure when increasing the string coupling back to the black hole regime. These glues are chosen exactly such that the back-reacted state still preserves the same number of global supersymmetries, but enhances the system locally to be 1/2-BPS.

This thesis aims at applying this concept of local supersymmetry enhancement to general supersymmetric systems. We determine the glues that enhance a general 1/4- or 1/8-BPS system to 16 preserved supersymmetries locally, without breaking any further supersymmetry globally. Together with my supervisor, we recently presented a list of all possible glues for 1/4-BPS systems in Type II string theory in \cite{Eckardt:2023nmn}. Furthermore, we use these glues as building blocks to derive the glues of the super-maze, and generalize it to other systems with three non-vanishing net charges, called main charges. We repeat these results in this thesis and expand on them. 

Additionally, there is another local supersymmetry enhancement structure we discuss in this thesis. It uses fewer glues than the super-maze, but also enhances the supersymmetry locally. The main difference between those two structures is that the super-maze exclusively uses internal glues; however, the other structure generally requires the glues to extend along at least one direction in the transverse space. This explicitly breaks the rotational symmetry of the black hole. The prime example for this kind of structure are the superstrata \cite{Bena:2011uw}. The main charges of the superstrata are a D-string, a $\D{5}$-brane, and momentum $\W$. 
This system is enhanced by including glues that extend along two different directions in the transverse space. The original configuration is smeared along this 2-dimensional sheet, resulting in an extended object of codimension 3. This configuration has now been studied for over a decade, but the entropy of the superstrata is parametrically smaller than the entropy of the $\D{1}$-$\D{5}$-$\W$ black hole \cite{Shigemori:2019orj}. 
We will also discuss the structure of the glues of the superstrata in this thesis, and generalize it to other systems with three main charges. 

We start by introducing the notion of local supersymmetry enhancement in chapter\,\ref{chap:LocalSupersymmetryEnhancement} from a supersymmetric point of view, and apply it to a general system with two main charges. This builds the foundation for an enhancement of multi-charge systems. It introduces the general equations the charge densities of the dipoles and the main branes have to satisfy, as well as showcases these for the simplest example. The techniques used to enhance 2-charge systems will be repeatedly used for systems with more charges.

In chapter \ref{chap:DualityMap}, we combine the 1/4-BPS configurations in Type-II string theory into a map and connect them by T- and S-dualities. We apply the conditions on the glues for a 1/4-BPS system derived in the previous chapter to list all possible glues for the respective configurations. These 2-charge bound states form the building blocks for the more complicated structures presented in the next two chapters. With this list of glues, the glues for 3-charge systems can be determined quickly.

In the remaining part of the thesis, we turn to 1/8-BPS systems. In chapter \ref{chap:TriangleAnsatz}, the glue structure of the super-maze is generalized to different charges. This enables us to determine internal dipole charges for arbitrary supersymmetric systems with three main charges. This procedure is demonstrated to find two possible sets of internal glues for a $\D{1}$-$\D{5}$-$\W$ brane system. Furthermore, we present a procedure of switching dipole charges with main charges to generate a bound state for a new set of main branes. A naive ansatz for a system with four main charges fails. Although the structure looks promising, the resulting equations have to be modified in some way to generate a solution. 

The glue structure for the superstrata is discussed in chapter \ref{chap:HangerAnsatz}. We derive the conditions on the glues for a general three-charge system and present the glues of the superstrata and a configuration of three orthogonal $\D{2}$-branes. Furthermore, we approach a generalization to systems with four main charges and apply it to the $\D{2}$-$\D{2}$-$\D{2}$-$\D{6}$ system. This uses two transverse directions, resulting in a two-dimensional sheet in the transverse spacetime. 

We conclude with a summary in chapter \ref{chap:Conclusion}. 
In Appendix \ref{app:chap:BraneInvolutions} we include the supersymmetry involutions for strings, branes and momenta in Type II string theory, and in Appendix\,\ref{app:chap:dualityMapsWithGlues} list all possible glues for the 1/4-BPS configurations in the duality map derived in chapter\,\ref{chap:DualityMap}. 

  \chapter{Local supersymmetry enhancement and 1/4 BPS systems}\label{chap:LocalSupersymmetryEnhancement}

This chapter introduces the notion of local supersymmetry enhancement and examines local supersymmetry for a general 2-charge configuration. The three- and four-charge systems discussed in chapters \ref{chap:TriangleAnsatz} and \ref{chap:HangerAnsatz} heavily rely not only on the techniques, but also the results of this chapter. 

We start by giving a short review of supersymmetry in the context of the projector associated with a brane or bound state. The notion of local supersymmetry enhancement arises by adding more (dipole) charges to the projector without breaking supersymmetry further. 
The general conditions for local supersymmetry enhancement for 1/4-BPS systems and the solution for the charge densities are discussed in section \ref{sec:LSE-2Charge}.
We conclude with two interesting symmetries for local supersymmetry enhancement.

\section{Revision on supersymmetry}\label{sec:RevisionOnSupersymmetry}

The amount of supersymmetry preserved by a brane\footnote{We denote all the objects of the theory by branes, e.g. \D{0}-particles, fundamental strings, solitonic branes or Kaluza-Klein-monopoles.} in Type II string theory or 11-dimensional supergravity is given by the kernel's dimension of the associated projector $\Pi := \frac{1}{2} (1 + P)$
\begin{equation}
    \Pi \, \epsilon = 0
\end{equation}
or equivalently by 
\begin{equation}\label{eq:supersymmetry-involution}
    P \, \epsilon = - \epsilon \, ,
\end{equation}
where $P$ is its associated involution (see \cite{Smith:2002wn} for a review).

When considering multiple brane species with involutions $P_i$, the projector is generalized to\footnote{The projector can be defined with a positive or a negative sign in front of the involutions. This sign and also the orientation of the branes can be absorbed into the coefficients $\alpha_i$.}
\begin{equation}\label{eq:MultiChargeProjectorDefinition}
    \Pi = \frac{1}{2} ( 1 + \sum_i \alpha_i P_i ) \, .
\end{equation}
The coefficients $\alpha_i$ describe the ratio of the brane charge density $Q_i$ to the total mass density $M$. They have to be chosen in a way such that $\Pi$ is still a projector
\begin{equation}
    \Pi^2 = \Pi \, .
\end{equation}
Inserting equation \eqref{eq:MultiChargeProjectorDefinition}, this can be rearranged into the two conditions 
\begin{align}
    1 &= \sum_i \alpha_i^2 \qquad \text{and} \\
    0 &= \sum_{\substack{i,j \\ i < j}} \alpha_i \alpha_j \, \{ P_i , P_j \} \, .
\end{align}

In general, the involutions are hermitian traceless product structures, i.e.
\begin{align}
    \Tr{P} = 0 && P^2 = 1 \, .
\end{align}
They depend on the background, the embedding map of the brane and its worldvolume fields (see \cite{Bergshoeff:1997kr}). We do not consider worldvolume fields and restrict ourselves to trivial backgrounds. In this case, the involutions for the branes in Type IIA and Type IIB string theory and M-theory are given by a product of Gamma matrices along the directions of the brane, and a Pauli matrix characteristic for the type of brane. They are listed in Appendix \ref{app:chap:BraneInvolutions}.

For one brane along a fixed direction, half of the supersymmetry is broken. The eigenvalues of an involution can only be $+1$ and $-1$. Since $P$ is also traceless, the sum of its eigenvalues vanishes. Equation \eqref{eq:supersymmetry-involution} then dictates, that exactly half of the supersymmetry is broken.

For two branes, we have to satisfy equation \eqref{eq:supersymmetry-involution} for both branes simultaneously.
\begin{align}
    P_1 \, \epsilon = -\epsilon && P_2 \, \epsilon = -\epsilon
\end{align}
Considering this as an eigenvalue problem, we can find a simultaneous eigenbasis only if they commute. Then the product $P_1 P_2$ is also an involution. Apart from the trivial case ($P_1 = P_2$), it is again a product of Gamma matrices and a Pauli matrix, and therefore traceless.
Inspecting the dimensions of eigenspaces of $P_1$, $P_2$ and $P_1 P_2$, there is only one possible combination: a quarter of the supersymmetry remains (see \cite{Smith:2002wn} for details).

\section{Local supersymmetry enhancement}

The idea of local supersymmetry enhancement is to add further charges, called glues, to the system and enhance it to 16 supersymmetries locally \cite{Bena:2022wpl}. We denote the involutions of those charges by the letter $Q$ and assume that the involutions of all main charges and glues are distinct. Otherwise, the ansatz and projector can be simplified. 
The projector of the whole system is
\begin{equation}
    \Pi = \frac{1}{2} \left( 1 + \sum_i \alpha_i P_i + \sum_j \beta_j Q_j \right) \, ,
\end{equation}
where the sum runs over all main charges, or glues respectively. Furthermore, we introduced the coefficients $\alpha_i$ for the main charges and $\beta_j$ for the glues. They describe the ratio of the charge density of its corresponding brane to the total mass of the bound state. 
As mentioned before, $\Pi$ is a projector. This results in the following two conditions on the coefficients,
\begin{align}
    1 &= \sum_i \alpha_i^2 + \sum_j \beta_j^2 \, , \label{eq:firstProjectorCondition} \\
    0 &= \sum_{\substack{i_1,i_2 \\ i_1 < i_2}} \alpha_{i_1} \alpha_{i_2} \{ P_{i_1} , P_{i_2} \}
    + \sum_{i,j} \alpha_{i} \beta_{j} \{ P_{i} , Q_{j} \}
    + \sum_{\substack{j_1,j_2 \\ j_1 < j_2}} \beta_{j_1} \beta_{j_2} \{ Q_{j_1} , Q_{j_2} \} \, . \label{eq:secondProjectorCondition}
\end{align}
The first condition is a dependent equation after deriving the other equations governing the system. The second condition poses the main challenge in finding systems with more than two main branes and glues. As the main branes commute to preserve some global supersymmetry, the terms in the first sum have to cancel with terms in the third sum. 
When considering more than two main charges, the second sum generally becomes non-trivial, and further terms appear in the third sum. We face and solve those challenges in chapters \ref{chap:TriangleAnsatz} and \ref{chap:HangerAnsatz}. 

The trace of the projector is
\begin{equation}
    \Tr{\Pi} = \frac{1}{2} \left( \Tr{1} + \sum_i \alpha_i \, \Tr{P_i} + \sum_j \beta_j \, \Tr{Q_j} \right) = \frac{1}{2} \Tr{1} \, ,
\end{equation}
showing that it is \textit{locally} 1/2-BPS. If the coefficients $\alpha_i$ and $\beta_j$ vary in space, less supersymmetry is preserved globally. 
The goal is to find a projector that retains the global supersymmetry preserved by the main branes.
A common way to ensure this is to rewrite the projector into
\begin{equation} \label{eq:globalsupersymmetrycondition}
    \Pi = \sum_i f_i(x) \, \Pi_i \, ,
\end{equation}
where $\Pi_i := 1/2 \, (1+P_i)$ is the projector associated to the main brane $P_i$\footnote{We identify each brane with its involution $P_i$ or $Q_j$. This makes the language less convoluted and easier to understand.} and $f_i(x)$ are matrix-valued functions. They usually take the form
\begin{equation}
    f_i(x) = \alpha_i (x) + \sum_{k} \beta_k Q_k P_i \, ,
\end{equation}
where the sum runs over some subset of all glues. We refer to these as the glues associated to the main brane $P_i$, and illustrate this by putting them next to each other in the diagrams in the following chapters. This association is not unique but has some arbitrariness to it (see section \ref{sec:SymmetriesOfMainBranesAndGlues-2Charge}).

By rewriting the projector in terms of the projectors of the main charges, 
the supercharges preserved globally are determined by
\begin{equation}
    \Pi \, \epsilon = \sum_i f_i(x) \, \Pi_i \, \epsilon = 0 \, .
\end{equation}
As $f_i(x)$ are in general independent functions, the equation reduces to
\begin{equation}
    \Pi_1 \, \epsilon = \Pi_2 \, \epsilon = \ldots = \Pi_i \, \epsilon = 0 \, ,
\end{equation}
which are the equations determining the global supercharges without the presence of further glue charges.

\section{1/4-BPS systems}\label{sec:LSE-2Charge}

In this section we apply the ideas of local supersymmetry enhancement to a system with two main charges $P_1$ and $P_2$. Together with a pair of glues $Q_1$ and $Q_2$, the system's projector becomes
\begin{equation}
    \Pi = \frac{1}{2} \left( 1 + \alpha_1 P_1 + \alpha_2 P_2 + \beta_1 Q_1 + \beta_2 Q_2 \right) \, .
\end{equation}
As already discussed in section \ref{sec:RevisionOnSupersymmetry}, the two main branes preserve 1/4 of the supersymmetry globally, if they commute and are not equal, $P_1 \neq P_2$. To preserve these global supersymmetries in presence of the glues, we make the ansatz\footnote{We suppress the dependence of the coefficients $\alpha_i$ and $\beta_j$ and the functions $f_i$ on space.}
\begin{subequations}\label{eq:fAnsatz2Charges}
\begin{align}
    f_1 &= \alpha_1 + \beta_1 Q_1 P_1 \, , \\
    f_2 &= \alpha_2 + \beta_2 Q_2 P_2 \, .
\end{align}
\end{subequations}
Inserted into the projector
\begin{align}
    \Pi &= f_1 \Pi_i + f_2 \Pi_2 = \frac{1}{2} \left( \alpha_1 + \alpha_2 + \alpha_1 P_1 + \alpha_2 P_2 + \beta_1 Q_1 + \beta_2 Q_2 + \beta_1 Q_1 P_1 + \beta_2 Q_2 P_2 \right) \, ,
\end{align}
this results in the following conditions on the coefficients
\begin{align}
    1 &= \alpha_1 + \alpha_2 \\
    0 &= \beta_1 + \eta \, \beta_2 \, . \label{eq:betaRelation-2Charge}
\end{align}
The first equation is the usual BPS equation. In the second equation, we introduced $\eta := P_1 Q_1 Q_2 P_2$. To satisfy it, $\eta$ has to be proportional to the identity matrix. We can multiply $\eta$ by $P_1$ from the left and the right to bring $P_1$ from the front to the back, or any other involution that is currently on the first position. This shows that $\eta$ is invariant under circular shifts. We call this the cyclic property, analogous to the cyclic property of the trace. Furthermore, taking the determinant of $\eta$ shows that $\eta$ is actually only a sign,
\begin{equation}
    \det \eta = \det P_1 \det Q_1 \det Q_2 \det P_2 = \pm 1 \, ,
\end{equation}
where we used that the determinant of an involution is $+1$ or $-1$.

\begin{figure}[htb]
    \centering
    \includegraphics[width = 0.4 \textwidth]{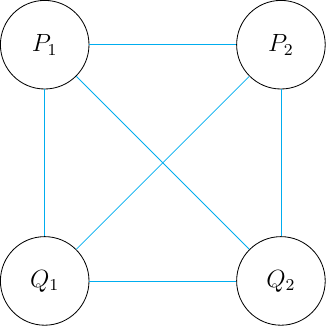}
    \hfill
    \includegraphics[width = 0.4 \textwidth]{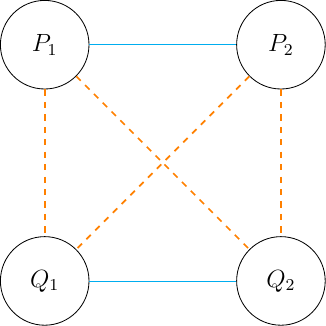}
    \caption{Illustration of the two possible sets of (anti-) commutation relations between two main branes and two glues. Blue lines denote a vanishing commutator, dotted orange lines a vanishing anticommutator. The left configuration is 1/2-BPS, while the right configuration is 1/4-BPS with 16 local supercharges.}
    \label{fig:AntiCommutationRelations2Charges}
\end{figure}

With these properties, we can derive most (anti-) commutation relations for the main branes and glues. From the cyclic property follows that either $P_1$ commutes with both $Q_1$ and $Q_2$, or it anticommutes with both
\begin{equation}\label{eq:CyclingP1ThroughEta}
    \eta = P_1 Q_1 Q_2 P_2 \overset{\text{cycl.}}{=} Q_1 Q_2 P_2 P_1 = Q_1 Q_2 P_1 P_2 \, .
\end{equation}
For both cases, we can compute the square of $\eta$ to prove that $Q_1$ and $Q_2$ commute
\begin{equation}\label{eq:SquaringEtaForCommutationRelations}
    1 = \eta^2 \overset{\text{cycl.}}{=} (P_1 Q_1 Q_2 P_2) \, (P_2 P_1 Q_1 Q_2) = Q_1 Q_2 \, Q_1 Q_2 \, .
\end{equation}
The remaining (anti-) commutators can be derived by cycling $Q_1$ and $Q_2$ through $\eta$. The subsequent relations are illustrated in figure \ref{fig:AntiCommutationRelations2Charges}.

With the (anti-) commutation relations between main branes and glues, we can return to the second projector condition \eqref{eq:secondProjectorCondition},
\begin{equation}
\begin{split} \label{eq:secondProjectorCondition2Charges}
    0 = & + \alpha_1 \alpha_2 \, \{ P_1 , P_2 \} + \beta_1 \beta_2 \{ Q_1 , Q_2 \} \\
    & + \alpha_1 \beta_1 \, \{ P_1 , Q_1 \} + \alpha_2 \beta_2 \{ P_2 , Q_2 \} \\
    & + \alpha_1 \beta_2 \, \{ P_1 , Q_2 \} + \alpha_2 \beta_1 \{ P_2 , Q_1 \} \, .
\end{split}
\end{equation}
For all commuting pairs of branes, we have to cancel the anticommutator in the equation above with some other non-vanishing anticommutator. As the main branes always commute, we illustrate this for them:
\begin{align}
    \alpha_1 \alpha_2 \, \{ P_1 , P_2 \} &\overset{\eta^2 = 1}{=} \eta \, \alpha_1 \alpha_2 \, (P_1 \eta P_2 + P_2 \eta P_1) \nonumber\\
    &\overset{\text{cycl.}}{=} \eta \, \alpha_1 \alpha_2 \, \left( P_1 (P_1 Q_1 Q_2 P_2) P_2 + P_2 (Q_2 P_2 P_1 Q_1) P_1 \right) \nonumber\\
    &= \eta \, \alpha_1 \alpha_2 \, (Q_1 Q_2 + Q_2 Q_1) = \eta \, \alpha_1 \alpha_2 \, \{ Q_1 , Q_2 \}\label{eq:cancellationanticommutators}
\end{align}
This introduces a new equation,
\begin{equation}
    \alpha_1 \alpha_2 = - \eta \, \beta_1 \beta_2 \, .
\end{equation}
We heavily rely on this computation in the coming chapters. As soon as we have a sign like $\eta$ (with the same (anti-) commutation relations between its constituents), we can introduce an equation for the corresponding coefficients to cancel the anticommutators that appear in the second projector condition.\footnote{More precisely, $P_1$ and $Q_1$ have to have the same (anti-) commutation relation as $P_2$ and $Q_2$.}

To summarize, we have derived the following independent equations,
\begin{subequations}
\begin{align}\label{eq:LSEEquations2Charges}
    1 &= \alpha_1 + \alpha_2 \\
    0 &= \beta_1 + \eta \, \beta_2 \\
    \alpha_1 \alpha_2 &= - \eta \, \beta_1 \beta_2 \, ,
\end{align}
\end{subequations}
with two extra equations for the left set of (anti-) commutation relations in figure \ref{fig:AntiCommutationRelations2Charges},
\begin{subequations}
\begin{align}
    \alpha_1 \beta_1 &= - \eta \, \alpha_2 \beta_2 \, \text{and} \\
    \alpha_1 \beta_2 &= - \eta \, \alpha_2 \beta_1 \, .
\end{align}
\end{subequations}

The added equations for the left set of (anti-) commutation relations, i.e. where all branes commute with each other, restrict the coefficients heavily. Solving this system of equations, results in a 1/2-BPS bound state, with $\alpha_i$ and $\beta_j$ are $\pm 1/2$. Without a continuous parameter that we can vary, there is no notion of local supersymmetry, and we can discard this case. 

However, for the right set of (anti-) commutation relation, i.e. where main branes and glues anticommute, the solution has a free parameter $\theta$,
\begin{subequations}\label{eq:solutionLSE2Charges}
\begin{align}
    \alpha_1 &= \cos^2(\theta) \\
    \alpha_2 &= \sin^2(\theta) \\
    \beta_1 &= \kappa \, \sin(\theta) \cos(\theta) \\
    \beta_2 &= - \eta \, \kappa \, \sin(\theta) \cos(\theta) \, .
\end{align}
\end{subequations}
$\kappa = \pm 1$ is a sign for the orientation of the glues. Here, we can also see the interpretation of the sign $\eta$. It controls the orientation of the glues with respect to each other.

There are two things we have to mention:
Firstly, with equations \eqref{eq:LSEEquations2Charges}, we can derive the first projector condition \eqref{eq:firstProjectorCondition} as announced in the previous chapter,
\begin{align}
    1 = 1^2 = (\alpha_1 + \alpha_2)^2 = \alpha_1^2 + \alpha_2^2 + 2 \, \alpha_1 \alpha_2 = \alpha_1^2 + \alpha_2^2 + \beta_1^2 + \beta_2^2 \, .
\end{align}
Secondly, the solution presented does not include negative values for $\alpha_1$ and/or $\alpha_2$. This is an artifact of defining the projectors with a positive sign instead of $\Pi_i = 1/2 \, ( 1 - P_i )$. Another way to generate negative solutions is to reverse the orientation in $P_1$ and/or $P_2$. This still produces the same solution for the coefficients (except for the sign $\eta$), which is equivalent to a negative solution for the original $P_1$ and/or $P_2$. This is also the case for the three- and four-charge systems considered in chapters \ref{chap:TriangleAnsatz} and \ref{chap:HangerAnsatz}.

\section{Symmetries of main branes and glues}\label{sec:SymmetriesOfMainBranesAndGlues-2Charge}

With the ansatz for $f_i$ in the equations \eqref{eq:fAnsatz2Charges}, we associated each glue with a main brane. This association is arbitrary in the sense that we would get the same solution if we switch the glues \cite{Li:2023jxb}. This is also reflected in the solution \eqref{eq:solutionLSE2Charges}. If we map $\kappa$ to $\kappa' = - \eta \, \kappa$, $\beta_1$ becomes $\beta_2$ and vice versa. Equivalently, the main branes can also be switched by the map $\theta \mapsto \theta' = \theta + \pi/2$ and $\kappa \mapsto \kappa' = - \kappa$.

There is another symmetry, but only for the existence of a local supersymmetry enhancement. For this, let us summarize its conditions from the previous section. \begin{subequations}
\begin{gather}
    \eta := P_1 Q_1 Q_2 P_2 = \pm 1 \\
    [P_1,P_2] = \{P_1,Q_1\} = 0
\end{gather}
\end{subequations}
Remember, all the other (anti-) commutation relations follow with the cyclic property of $\eta$. These conditions are invariant under exchanging all main branes with all glues, $P_i \mapsto Q_i$ and vice versa. The (anti-) commutation relations are trivial. The invariance of $\eta$ follows from its cyclic property,
\begin{align}
    \eta' = Q_1 P_1 P_2 Q_2 \overset{\text{cycl.}}{=} P_1 P_2 Q_2 Q_1 \overset{\text{comm.}}{=} P_2 P_1 Q_1 Q_2 \overset{\text{cycl.}}{=} P_1 Q_1 Q_2 P_2 \, .
\end{align}
The resulting configuration, i.e. with main branes $Q_i$ and glues $P_j$, is different to the original one. 
Only the existence of a local supersymmetry enhancement follows from the existence of one for main branes $P_i$ and glues $Q_j$.

  \chapter{The duality map}\label{chap:DualityMap}

In the upcoming chapters, we construct multi-charge configurations by glueing main charges together with a pair of glues. To derive specific configurations, it is beneficial to know all possible glues for a pair of main branes. In this chapter, we first create a map of 1/4-BPS configurations in Type II string theory, and secondly append a list of possible glues to every configuration. At the end, we highlight common and important glues found in the duality map.

\section{Finding 1/4-BPS configurations}\label{sec:DualityMap:1/4BPSconfigurations}

As discussed in the previous chapter, two distinct branes preserve a quarter of the original supersymmetry, if their involutions commute. The involutions of the branes in Type II string theory are listed in Appendix \ref{app:chap:BraneInvolutions}. Implementing those in Mathematica, we can create a list of all possible commuting pairs of branes. For branes $A$ and $B$ with $m$ common directions, we denote such a pair as
\begin{align}\label{eq:Notation-2Charge-Orthogonal}
    A \perp B \: (m) \, ,
\end{align}
or, if one brane is completely inside or parallel to the other,
\begin{align}\label{eq:Notation-2Charge-Parallel}
    A \parallel B \: (m) \, .
\end{align}

\begin{sidewaysfigure}[tb]
    \centering
    \includegraphics[width = \textwidth]{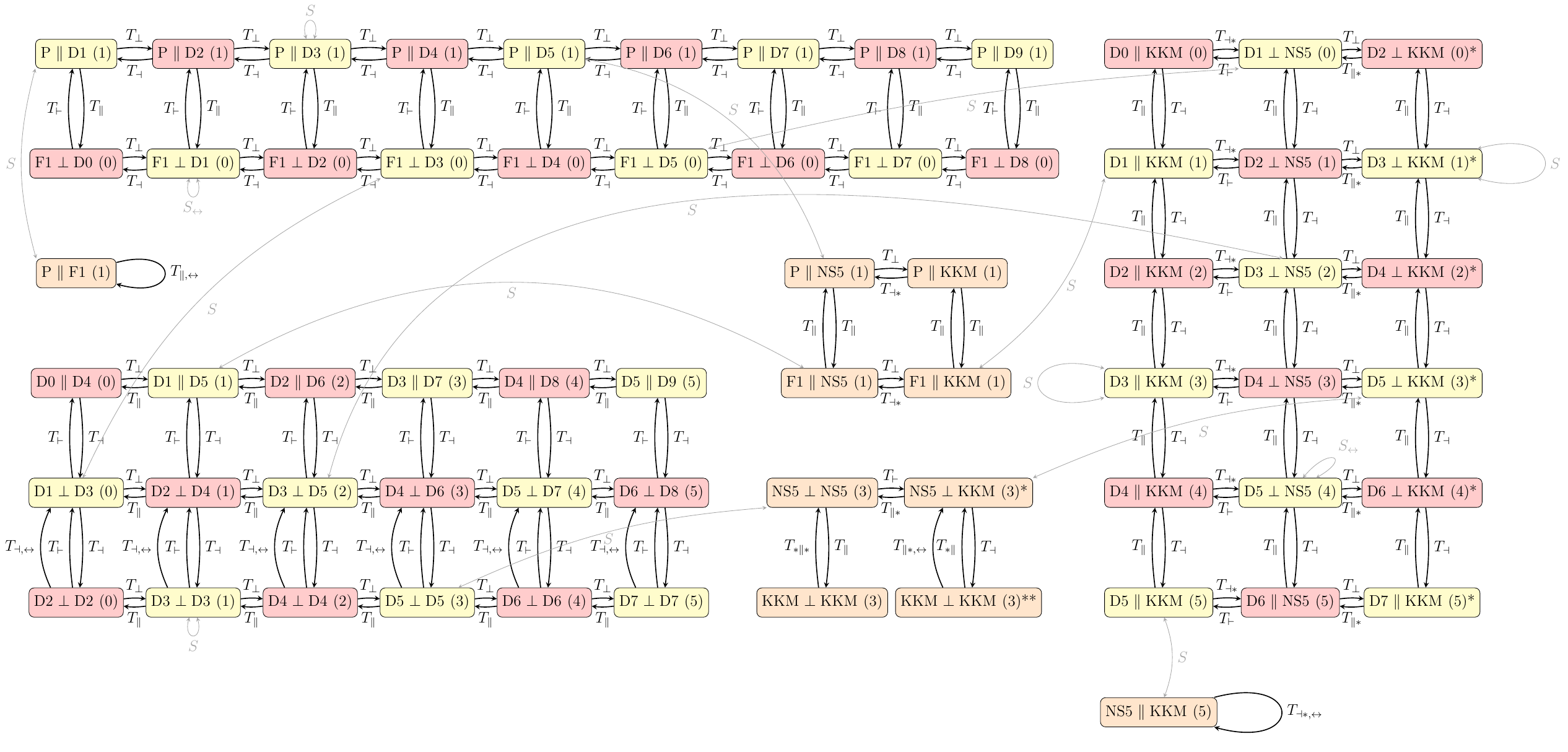}
    \caption{Duality map for standard 1/4-BPS configurations. For details or notation we refer to the text in section \ref{sec:Notation}.}
    \label{fig:DualityMap}
\end{sidewaysfigure}

\begin{figure}[tb]
    \centering
    \includegraphics[width = \textwidth]{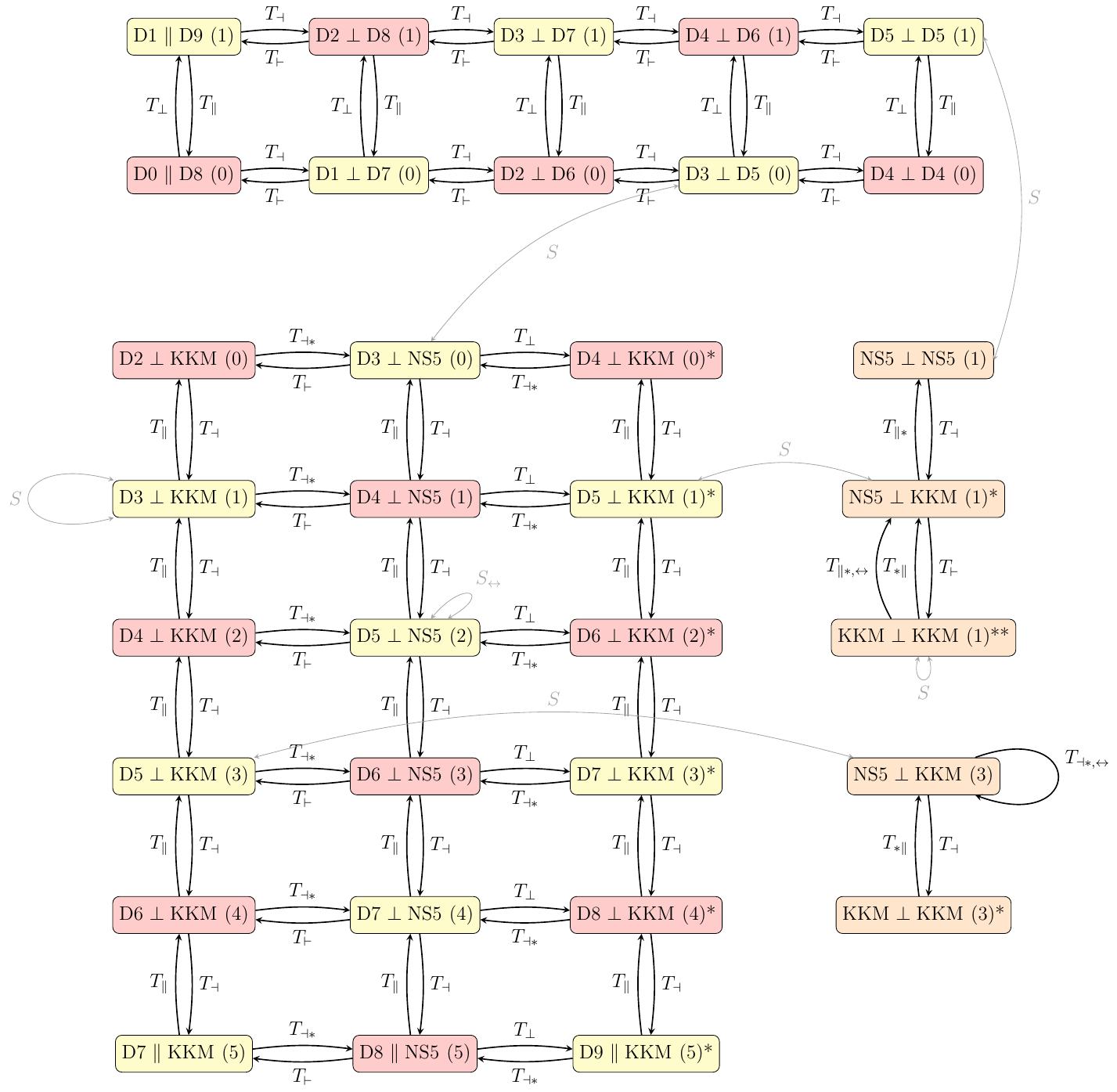}
    \caption{Duality map for non-standard 1/4-BPS configurations. For details or notation we refer to the text in section \ref{sec:Notation}.}
    \label{fig:DualityMapNonStandard}
\end{figure}

Many of the these configurations are T- and S-dual to each other. This creates an intermingled web of 1/4-BPS configurations. We have two completely disconnected components. The first one, called the \textit{standard} configurations, is displayed in figure \ref{fig:DualityMap}, the second one, the \textit{non-standard} configurations, in figure \ref{fig:DualityMapNonStandard}. We grouped the configurations into blocks closed under T-dualities (black arrows). These blocks are then connected to other blocks via S-dualities (gray arrows). 

\clearpage
\section{Adding glues to the duality map}

The next step is to find the possible local supersymmetry enhancements for the configurations in the duality map by applying the conditions from the previous chapter. These conditions are
\begin{subequations}
\begin{gather}
    \eta := P_1 Q_1 Q_2 P_2 = \pm 1 \quad \text{and} \\
    [P_1,P_2] = \{P_1,Q_1\} = 0 \, .
\end{gather}
\end{subequations}
As the main glues $P_1$ and $P_2$ are fixed and commuting, we determine candidates for the glues $Q_1$ and $Q_2$ by inspecting the first condition, $\eta = \pm 1$. Afterwards, we check whether the main branes anticommute with the glues.

The involutions are a tensor product of two parts, the Gamma matrices and Pauli matrix. The Gamma matrices describe which dimensions the brane is extended along. The Pauli matrix encodes the type of the brane. In order to satisfy $\eta = \pm 1$, both parts have to be proportional to the identity matrix. We illustrate our process of finding all possible glues for the configuration $\W[1] \parallel \D[12]{2} \text{ (1)}$. The condition on $\eta$ then becomes $Q_1 \, Q_2 = \Gamma^{2} \, \sigma_1$. This equation dictates that the dimensions of the two glues have a difference of $1$. Furthermore, we can inspect the algebra of Pauli matrices, $\sigma_1 = 1 \cdot \sigma_1 = - i \sigma_2 \cdot \sigma_3$. This leaves seven possible glues (apart from coordinate transformations),
\begin{table}[h]
    \centering
    \begin{tabular}{c}
        $\W[1] \parallel \D[12]{2}$ \\
        $\mathbf{{P}_9 \parallel D{2}_{29}}$ \\
        $\NS[16789] \parallel \D[126789]{6}$ \\
        $\mathbf{ NS5_{56789} \parallel D{6}_{256789} }$ \\
        $\mathbf{ F{1}_{2} \parallel D{0} }$ \\
        $\KKM[12789] \parallel \D[1789]{4}$ \\
        $\mathbf{ KKM_{26789} \parallel D{4}_{6789} }$ \, .
    \end{tabular}
\end{table}

The first one can be discarded immediately, as it violates the assumption that all main branes and glues have to be distinct. Computing $\{P_1,Q_1\}$ for the remaining six pairs of glues, it vanishes for the glues highlighted by a bold font. There are four distinct pairs of glues that enhance a $\D{2}$ brane with momentum to 16 local supercharges.

\begin{figure}[htb]
    \centering
    \includegraphics[width = 0.8 \textwidth]{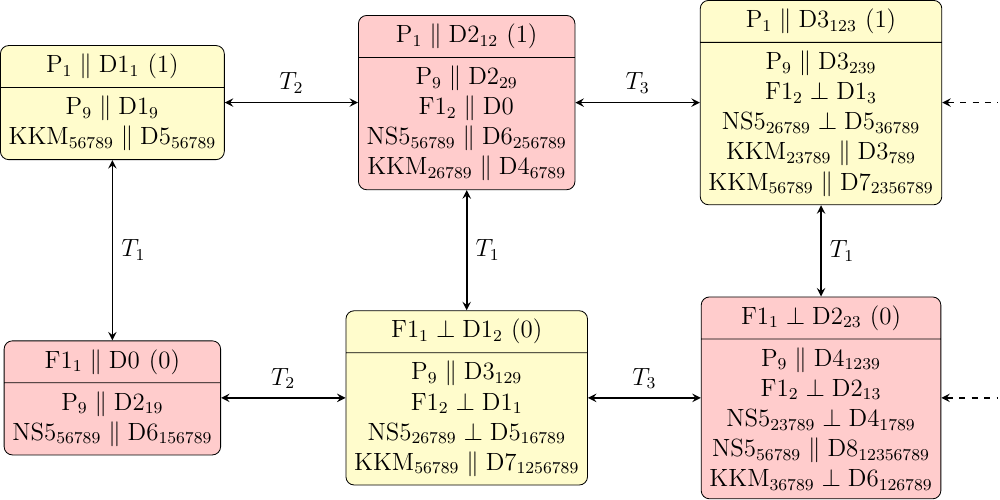}
    \caption{A part of the duality map with possible local supersymmetry enhancements. For details, we refer to the appendix \ref{sec:Notation}.}
    \label{fig:P-Dp-Configurations-Shortened}
\end{figure}

We repeated this procedure for all 1/4-BPS configurations in the duality map and appended all possible glues to the nodes. The resulting diagram is huge. We split it into the blocks closed under T-dualities, and included them in appendix \ref{app:chap:dualityMapsWithGlues}. A fragment of the $\W$-$\D{p}$ and $\F$-$\D{q}$ standard configuration can be seen in figure \ref{fig:P-Dp-Configurations-Shortened}. In section \ref{sec:SymmetriesOfMainBranesAndGlues-2Charge} we discussed a symmetry under exchanging main branes with glues. This symmetry is also reflected in the diagram: 
The pair $\F[2] \parallel \D{0}$ enhances $\W[1] \parallel \D[12]{2} \text{ (1)}$ to 16 local supercharges. The 'inverse' configuration, $\W[9] \parallel \D[19]{2}$ enhancing $\F[1] \parallel \D{0} \text{ (0)}$, is also represented in figure \ref{fig:P-Dp-Configurations-Shortened}.

\section{Discussion of important configurations}\label{sec:ImportantConfigurations-2Charge}

That the dual to a local supersymmetry enhanced system is also enhanced, explains the myriad of possible glues. In this section, we try to illuminate some similarities between configurations in this web and highlight important ones. For this, we denote an enhanced system with main branes $A$ and $B$ and glues $C$ and $D$ by
\begin{equation}
    [ \, A \perp B \; (m) \, ] \;
    \rightarrow \;
    ( \, C \perp D \, ) \, .
\end{equation}

The first configuration is an oscillating D- or F-string with momentum,
\begin{equation}
    [ \, \D[y]{1} \parallel \W[y] \; (1) \, ] \;
    \rightarrow \;
    ( \, \D[\psi]{1} \parallel \W[\psi] \, ) \, .
\end{equation}
By tuning the parameter $\theta(x)$, we can create a profile of the string in the $(y,\psi)$ plane. Locally, we only have a section of a string along a direction $\hat{y}$ that is tilted by an angle $\theta$ with respect to $y$, and a transverse momentum carried by the string. This small section preserves 16 supercharges. Varying the angle $\theta$ in space, however, changes 8 of the 16 preserved local supercharges, giving a configuration with 8 global supersymmetries. 
A supergravity description can be found in \cite{Dabholkar:1995nc}.
T-dualizing the D-string along $p$ other directions does not change the momenta. The result is an oscillating $\D{(p+1)}$-brane,
\begin{equation}
    [ \, \D[y1 \dotsc p]{(p+1)} \parallel \W[y] \; (1) \, ] \;
    \rightarrow \;
    ( \, \D[\psi 1 \dotsc p]{(p+1)} \parallel \W[\psi] \, ) \, .
\end{equation}

Another important configuration that can also be found in figure \ref{fig:P-Dp-Configurations-Shortened} is the supertube \cite{Mateos:2001qs,Emparan:2001ux},
\begin{equation}
    [ \, \F[y] \parallel \D{0} \; (0) \, ] \;
    \rightarrow \;
    ( \, \D[y\psi]{2} \parallel \W[\psi] \, ) \, .
\end{equation}
A tubular $\D{2}$ brane is prevented from collapse by an angular momentum. The angular momentum is generated by a uniform electric field, the fundamental string, and magnetic flux, the $\D{0}$ charge. This gives a superstring with constant $\D{0}$ charge smeared along the circular direction of the $\D{2}$-supertube. Again, this has many T-dual analogues,
\begin{equation}
    [ \, \F[y] \perp \D[1 \ldots p]{p} \; (0) \, ] \;
    \rightarrow \;
    ( \, \D[y\psi 1 \ldots p]{(p+2)} \parallel \W[\psi] \, ) \, .
\end{equation}

Lastly, we have the following combination of fundamental strings and D-strings,
\begin{equation}
    [ \, \F[y] \perp \D[z]{1} \; (0) \, ] \;
    \rightarrow \;
    ( \, \F[z] \perp \D[y]{1} \, ) \, .
\end{equation}
The authors of \cite{Hashimoto:2003pu} interpret this as two $(p,1)$-strings intersecting at an angle determined by $p$. The bisecting line of the two D-strings is along $z$, while its perpendicular is along $y$. The net charge of the D-strings along $y$ vanishes, as the two D-strings cancel each other in this direction. On the other hand, the electric fluxes on the D-string, i.e. the fundamental string fluxes, point towards $y$ and cancel each other along $z$. 
A supersymmetry preserving deformation can separate the two strings by the distance $a$ forming a hyperbolic profile
\begin{equation}
    \lambda (x) = \pm \sqrt{p^2 x^2 + a^2} \, .
\end{equation}
For a great illustration, see figures 1 to 4 in \cite{Hashimoto:2003pu}.

A similar configuration is the Callan-Maldacena spike, or BIon spike \cite{Callan:1997kz}. 
It describes a D-string ending on a $\D{3}$-brane, but instead of a sharp transition at the end point of the string, a smooth transition between the two objects occurs. Again, they form a hyperbolic profile in the $(y,r)$ plane, where $y$ is the direction of the D-string and $r$ the radial direction of the \D{3} brane. The $\D{1}$ and $\D{3}$ flux both follow this profile \cite{Hashimoto:2003pu,Callan:1997kz}. For a finite D-string stretched between a pair of $\D{3}$-branes, the shape between the branes is modified to an X-like cross-section \cite{Hashimoto:1997px}.
In our notation, both systems correspond to the configuration
\begin{equation}
    [ \, \D[y]{1} \perp \D[r \theta \phi]{3} \; (0) \, ] \;
    \rightarrow \;
    ( \, \D[r]{1} \perp \D[y \theta \phi]{3} \, ) \, .
\end{equation}

For all the previous configurations, we can T-dualize to get new configurations that can be interpreted similarly. This gives us an intuitive explanation of many low dimensional, local supersymmetry enhanced systems in the duality web. Additionally, there are many more configurations with glues which extend along additional dimensions. We have already given one example of this phenomenon, the supertube. Other instances are 
\begin{equation}
    [ \, \D[y]{1} \perp \D[y 1234]{5} \; (1) \, ] \;
    \rightarrow \;
    ( \, \KKM[1234 \psi] \perp \W[\psi] \, ) 
\end{equation}
or
\begin{equation}
    [ \, \W[y] \perp \D[y]{1} \; (1) \, ] \;
    \rightarrow \;
    ( \, \KKM[12345] \perp \NS[12345] \, ) \, .
\end{equation}
They can be interpreted as a point-like structure in the transverse space $\mathbb{R}^4$, i.e. the non-compact dimensions, that is 'puffed up' into a closed curve (or higher-dimensional sheet) along $\psi$ for the former configuration, and $x^1$ to $x^5$ for the latter.

We want to highlight two further configurations, whose glues are inside the main branes,
but whose interpretation is not clear.
\begin{equation}
    [ \, \D{0} \perp \D[1234]{4} \; (0) \, ] \;
    \rightarrow \;
    ( \, \D[12]{2} \perp \D[34]{2} \, ) 
\end{equation}
seems to be \D{2} dipole charges resolved in a \D{4} background with constant $\D{0}$ charge.

The second configuration is
\begin{equation}
    [ \, \F[y] \perp \D[12345]{5} \; (0) \, ] \;
    \rightarrow \;
    ( \, \D[y]{1} \perp \NS[12345] \, ) \, .
\end{equation}
It stands out as its glues do not use extra dimensions. It could be related to $(p,q)$-strings and 5-branes. Further research will hopefully explain these configurations.

  \chapter{The triangle ansatz}\label{chap:TriangleAnsatz}

In the last chapters, we enhanced the 8 supersymmetries of a 2-charge system locally to 16 supersymmetries by connecting the charges with glues, and derived the glues for configuraitons in Type II string theories. This yielded many known configurations like the supertube or the Callan-Maldacena spike. The next step is to generalize this to more charges. The naive ansatz is to connect all main charges with each other with a pair of glues that enhances the 2-charge subsystem to 16 local supersymmetries. For three charges, this gives us the \textit{triangle ansatz}. 

Again, we derive the general conditions on a system of this form and find the general solution for the charge densities in section \ref{sec:SolutionTriangleAnsatz}. In sections \ref{subsec:Finding3ChargeConfigurations} and \ref{sec:3ChargeSymmetries} we present a procedure for finding the glues for a three-charge system, and discuss a symmetry on the existence of local supersymmetry enhancement, similar to the symmetry for the 2-charge case discussed in \ref{sec:SymmetriesOfMainBranesAndGlues-2Charge}. A few examples of 1/8-BPS configurations with 16 local supercharges are also presented in these sections.
Lastly, we show that this naive ansatz no longer works for 4-charges.

\section{Solution to the triangle ansatz}\label{sec:SolutionTriangleAnsatz}

\begin{figure}[htb]
    \centering
    \includegraphics[width = 0.4 \textwidth]{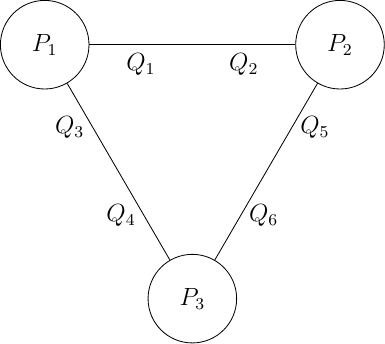}
    \caption{Triangle ansatz of the local supersymmetry enhancement (LSE) for a three-charge, $1/8$-BPS system. The involutions of the main branes are denoted by $P_i$, while the involutions of the glues are denoted by $Q_j$. The glues on the line between two main branes are involved in the two-charge LSE of those two main branes.}
    \label{fig:TriangleAnsatz}
\end{figure}

The starting point is a system with three main charges with involutions $P_1$, $P_2$ and $P_3$. These are distinct and commuting such that they form an 1/8-BPS system. Recalling the ansatz for a 1/4-BPS system, the main issue with further charges are the anticommutators appearing in the projector condition. At this point, there are three non-vanishing anticommutators that have to cancel with others. The most straightforward ansatz is to use three pairs of glues, $Q_1$ to $Q_6$. This ansatz is depicted in figure \ref{fig:TriangleAnsatz}. Its projector is
\begin{equation}
    \Pi = \frac{1}{2} \left( 1 + \alpha_1 (x) \, P_{1} + \alpha_2 (x) \, P_{2} + \alpha_3 (x) \, P_{3} 
    + \sum_{j=1}^{6} \beta_j (x) \, Q_{j} \right) \, .
\end{equation}
We again suppress the dependence of the coefficients $\alpha_i$ and $\beta_j$ on space.

To preserve 1/8 of the supersymmetry globally, we make the ansatz analogoues to the one in chapter \ref{chap:LocalSupersymmetryEnhancement},
\begin{subequations}
\begin{align}
    f_1 &= \alpha_1 + \beta_1 Q_1 P_1 + \beta_3 Q_3 P_1 \\
    f_2 &= \alpha_2 + \beta_2 Q_2 P_2 + \beta_5 Q_5 P_2 \\
    f_3 &= \alpha_3 + \beta_4 Q_4 P_3 + \beta_6 Q_6 P_3 \, .
\end{align}
\end{subequations}
This gives us similar equations to the 2-charge ansatz,
\begin{subequations}\label{eq:globalSupersymmetryCondition-3Charges}
\begin{align}
    1 &= \alpha_1 + \alpha_2 + \alpha_3 \label{eq:BPSCondition3Charges}\\
    \beta_1 &= - \beta_2 \, \eta_{12} \label{eq:pairOfGlues12}\\
    \beta_3 &= - \beta_4 \, \eta_{13} \label{eq:pairOfGlues34}\\
    \beta_5 &= - \beta_6 \, \eta_{23} \label{eq:pairOfGlues56}
\end{align}
\end{subequations}
where we introduced the signs $\eta_{12} = P_1 Q_1 Q_2 P_2$, $\eta_{13} = P_1 Q_3 Q_4 P_3$ and $\eta_{23} = P_2 Q_5 Q_6 P_3$. Their indices are the indices of its involved main branes. By the same reasoning as in section \ref{sec:LSE-2Charge}, these are signs with a cyclic property and its constituents follow the same (anti-) commutation relations.

To cancel the anticommutators of the main branes with those of its glues, we introduce the following three equations,
\begin{subequations}\label{eq:quadraticEquations}
\begin{align}
    \alpha_1 \alpha_2 &= - \eta_{12} \, \beta_1 \beta_2 \label{eq:quadraticEquation12}\\
    \alpha_1 \alpha_3 &= - \eta_{13} \, \beta_3 \beta_4 \label{eq:quadraticEquation13}\\
    \alpha_2 \alpha_3 &= - \eta_{23} \, \beta_5 \beta_6 \, . \label{eq:quadraticEquation23}
\end{align}
\end{subequations}
These equations also follow from the first projector condition \eqref{eq:firstProjectorCondition}.

There remains only one condition that our system has to satisfy, the second projector condition \eqref{eq:secondProjectorCondition}. Up to now, we have not discussed the anticommutator between a main brane and the glues on the opposite side of the triangle, see \ref{fig:TriangleAnsatz}, and between two glues on different sides of the triangle.
To assume that all anticommutators vanish constrains the main branes and glues quite heavily. Instead, we cancel all anticommutators with main branes with others. To determine their partners, we make the following observation:
\begin{align*}
    (\alpha_1 \alpha_2) \, (\beta_3 \beta_5) &\sim (\beta_1 \beta_2) \, (\beta_3 \beta_5) \\
    \sim (\alpha_1 \beta_5) \, (\alpha_2 \beta_3) &\sim (\beta_2 \beta_3) \, (\beta_1 \beta_5) \, .
\end{align*}
We deliberately ignored signs in this equation, indicated by $\sim$.
The important point is that we can split each line into two equations, indicated by the brackets. The first line is equation \eqref{eq:quadraticEquation12} and an identity. The second equation is split into two non-trivial equations that would follow from cancelling two anticommutators. Repeating this for the other equations gives us six equations,
\begin{subequations}\label{eq:mixedQuadraticEquations}
\begin{align}
    \alpha_1 \beta_5 &= - \mu_{15} \, \beta_2 \beta_3 \label{eq:mu15equation}\\
    \alpha_1 \beta_6 &= - \mu_{16} \, \beta_1 \beta_4 \\
    \alpha_2 \beta_3 &= - \mu_{23} \, \beta_1 \beta_5 \label{eq:mu23equation}\\
    \alpha_2 \beta_4 &= - \mu_{24} \, \beta_2 \beta_6 \\
    \alpha_3 \beta_1 &= - \mu_{31} \, \beta_3 \beta_6 \label{eq:mu31equation}\\
    \alpha_3 \beta_2 &= - \mu_{32} \, \beta_4 \beta_5 \, . \label{eq:mu32equation}
\end{align}
\end{subequations}
The signs $\mu_{ij}$ are defined as follows. For clarification, we use $\mu_{15}$ as an example. The first index denotes the main brane, $P_1$ in this case, and the second one a glue on the other side of the triangle, $Q_5$. In between those two involutions are the involutions of the glues with the same anticommutator (up to a sign). We assume those to be the glue next to $Q_5$ but on another side of the triangle, $Q_2$, and the glue next to $P_1$ on the last side of the triangle, $Q_3$.\footnote{As discussed in section \ref{sec:SymmetriesOfMainBranesAndGlues-2Charge}, we have the freedom of switching two glues on the same side of the triangle. We have chosen this convention to easily derive the signs and equations by following the presented scheme. Switching two glues breaks this scheme, but is allowed.} This gives $\mu_{15} = P_1 (Q_2 Q_3) Q_5$. These glues can also be read off of the equations \eqref{eq:mixedQuadraticEquations} above.

For the same reasons as for $\eta$, $\mu_{i,j}$ is a sign with a cyclic property. Furthermore, we already know that its main glue $P_i$ anticommutes with the other two glues inside $\mu_{i,j}$ (not $Q_j$). Due to the cyclic property, $P_i$ then has to commute with $Q_j$. Therefore, the (anti-) commutation relations have the same structure as for $\eta$.

Now, we can check that equations \eqref{eq:quadraticEquations} follow from equations \eqref{eq:mixedQuadraticEquations}, 
\begin{equation}
\begin{split}
    \alpha_1 \beta_5 \, \alpha_2 \beta_3 &= \mu_{15} \mu_{23} \, \beta_2 \beta_3 \, \beta_1 \beta_5 \\
    &= (P_1 Q_2 Q_3 Q_5 \, P_2 Q_1 Q_5 Q_3) \, \beta_2 \beta_3 \, \beta_1 \beta_5 \\
    &= - (P_1 Q_1 Q_2 P_2) \, \beta_2 \beta_3 \, \beta_1 \beta_5 \\
    &= - \eta_{12} \, \beta_1 \beta_2 \, \beta_3 \beta_5
\end{split}
\end{equation}
With equations \eqref{eq:globalSupersymmetryCondition-3Charges}, equations \eqref{eq:mixedQuadraticEquations} can further be reduced to three equations. The second projector condition is satisfied, as every pair of involutions now appears in one of the signs $\eta_{ij}$ or $\mu_{ij}$. The system is governed by the following seven equations:
\begin{subequations}
\begin{align}
    1 &= \alpha_1 + \alpha_2 + \alpha_3 \, , \tag{\eqref{eq:BPSCondition3Charges}}\\
    \beta_1 &= - \beta_2 \, \eta_{12} \, , \tag{\eqref{eq:pairOfGlues12}}\\
    \beta_3 &= - \beta_4 \, \eta_{13} \, , \tag{\eqref{eq:pairOfGlues34}}\\
    \beta_5 &= - \beta_6 \, \eta_{23} \, , \tag{\eqref{eq:pairOfGlues56}}\\
    \alpha_1 \beta_5 &= - \mu_{15} \, \beta_2 \beta_3 \, , \tag{\eqref{eq:mu15equation}}\\
    \alpha_2 \beta_3 &= - \mu_{23} \, \beta_1 \beta_5 \, , \tag{\eqref{eq:mu23equation}}\\
    \alpha_3 \beta_1 &= - \mu_{31} \, \beta_3 \beta_6 \, . \tag{\eqref{eq:mu31equation}}
\end{align}
\end{subequations}
With the ansatz, $\alpha_1 = a^2$, $\alpha_2 = b^2$ and $\alpha_3 = c^2$, the BPS-equation becomes the equation of a 2-sphere. We can freely distribute the energy density among the main branes. Furthermore, we have two parameters to vary in space. The coefficients for the glues are fixed by $a$, $b$ and $c$ up to the orientation of the glues,
\begin{subequations}\label{eq:Solution-TriangleAnsatz}
\begin{align}
    \beta_1 &= \kappa_1 \, a b \\
    \beta_3 &= \kappa_3 \, a c \\
    \beta_5 &= \kappa_5 \, b c \\
    \beta_2 &= - \eta_{12} \, \beta_1 \\
    \beta_4 &= - \eta_{13} \, \beta_3 \\
    \beta_6 &= - \eta_{23} \, \beta_5 \, ,
\end{align}
\end{subequations}
with signs $\kappa_i = \pm 1$ satisfying $\kappa_1 \kappa_3 \kappa_5 = - \mu_{23}$.

\section{A procedure for finding 3-charge configurations}\label{subsec:Finding3ChargeConfigurations}

There are various assumptions in the previous chapter. On top of the (anti-) commutation relations, there are nine products of involutions that have to be signs to enhance the local supersymmetry of a 3-charge system. These are not independent,
\begin{subequations}\label{eq:relationsBetweenEtaAndMu}
\begin{align}
    \mu_{15} \mu_{16} &\sim - \eta_{12} \eta_{23} \eta_{13} \\
    \mu_{15} \eta_{12} &\sim - \mu_{23} \\
    \mu_{16} \eta_{12} &\sim - \mu_{24} \\
    \mu_{15} \eta_{13} &\sim - \mu_{32} \\
    \mu_{16} \eta_{13} &\sim - \mu_{31} \, 
\end{align}
\end{subequations}
where $\sim$ denotes equality up to a sign. This implies that we only need to check for four signs, $\eta_{12}$, $\eta_{13}$, $\mu_{15}$ and $\mu_{16}$, while the others follow. This is true regardless of the specific (anti-) commutation relations.\footnote{We assume that two involutions either commute or anticommute.} For the specific relations derived in the previous section, the equations above are true equalities.

Regarding the (anti-) commutation relations, we have already proved above that the (anti-) commutation relations of constituents of some $\mu_{ij}$ follow from the relation of the $\eta$s. A minimal set of relation, from which all other follow, is therefore given by
\begin{equation}
    \{ P_1 , Q_1 \} = \{ P_1 , Q_3 \} = \{ P_2 , Q_5 \} = 0 \, .
\end{equation}
The last anticommutator, $\{P_2,Q_5\}$ can be replaced by $\{Q_1,Q_3\}$ or $[P_2,Q_3]$. 
This can be seen when considering the constituents of $\mu_{23} = P_2 Q_1 Q_5 Q_3$. As showcased for $\eta$ in equations \eqref{eq:CyclingP1ThroughEta} and \eqref{eq:SquaringEtaForCommutationRelations}, we can derive all (anti-) commutation relations for a sign if we know the relations of one constituent with two of the other constituents. In this case, we know $\{P_2,Q_1\} = 0$ from $\{P_1,Q_1\}=0$. The second relation can be $\{ P_2 , Q_5 \} = 0$, but we can also use e.g. $\{Q_1,Q_3\}=0$ or $[P_2,Q_3]=0$. 
This is useful for the next step, a scheme on finding the glues for a given 3-charge system.
        
    
Suppose we have three (commuting) main branes, $P_1$, $P_2$ and $P_3$, in a Type-II string theory. The idea of the triangle ansatz is to have a 2-charge local supersymmetry enhancement for every pair of main branes. As we have listed all glues in the previous chapter, we can use figures \ref{fig:P-Dp-Configurations} to \ref{fig:Dp-NS5-NonStandard-Configurations-2} in the appendix to determine glues $Q_1$ to $Q_4$. This way, $\eta_{12}$ and $\eta_{13}$ are signs, and their constituents already obey the correct (anti-) commutation relations. 
For the correct (anti-) commutation relations, $P_1$ has to anticommute with $P_3$ or $P_4$. If it anticommutes with $P_4$, we switch and relabel the glues $P_3$ and $P_4$.\footnote{This ensures that we remain in the labelling scheme chosen in this thesis.} If $Q_1$ commutes with both, we have to choose different glues and start over. 
There remain two conditions to ensure local supersymmetry enhancement, $\mu_{15} = \pm 1$ and $\mu_{16} = \pm 1$. We can choose the two remaining glues as $Q_5 = P_1 Q_2 Q_3$ and $Q_6 = P_1 Q_1 Q_4$ such that these are always fulfilled.\footnote{The sign of $Q_5$ and $Q_6$, and equivalently the sign of $\mu_{15}$ and $\mu_{16}$ does not matter, as ultimately this is controlled by the sign of $\beta_5$ and $\beta_6$.} 
We label $Q_3$ as the glue that anticommutes with $Q_1$, and $Q_4$ with $Q_2$. 
But be aware, $Q_5$ and $Q_6$ have to correspond to objects in the theory!

For clarification, we us illustrate this procedure at the example of a $\D[y]{1}$-string  on a circle $S_y^1$, a $\D[y1234]{5}$-brane also wrapping the circle and a $T^4$ labelled by $1$, $2$, $3$ and $4$, and a momentum charge $\W[y]$ on the same $S_y^1$-circle. In \cite{Bena:2011uw}, a set of glues was already discovered using two degrees of freedom of the orthogonal $\mathbb{R}^4$.\footnote{The configuration presented in \cite{Bena:2011uw} does not fit within the triangle ansatz, but what we call the hanger ansatz. Those systems are covered in the next chapter.} We derive two different configurations that only use internal glues, one of them presented in \cite{Eckardt:2023nmn}. For the $\D[y]{1}$-$\W[y]$ subsystem, there is only one internal pair of glues, $\D[1]{1}$-$\W[1]$. This corresponds to an oscillating string inside the $\D{5}$-brane. Its direction is chosen arbitrarily as $x^1$. For the second subsystem, $\D{5}$-$\W$, there is again only one internal pair of glues, $\F$-$\D{3}$, but we have two choices for their directions. The $\F$ dipole charges can either be along the direction of oscillation of the $\D{1}$-string, or not, while the $\D{3}$ occupies the remaining directions of the torus.

\begin{figure}[h]
    \centering
    \begin{subfigure}{0.45\textwidth}
        \includegraphics{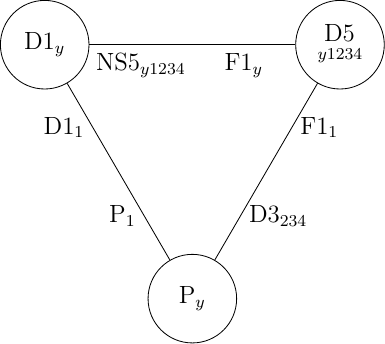}
    \end{subfigure}
    \begin{subfigure}{0.45\textwidth}
        \includegraphics{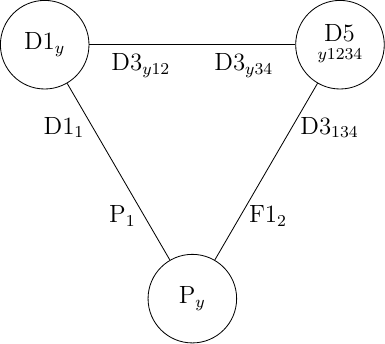}
    \end{subfigure}
    \caption{Graphical representation of two possible local supersymmetry enhancement of the $\D{1}$-$\D{5}$-$\W$ configuration with its respective glues.}
    \label{fig:D1-D5-P-Triangle}
\end{figure}

We focus on the second case. The glues are then $\F[2]$ and $\D[134]{3}$. With the natural choice to associate $\W[1]$ with $\W[y]$, the other glue next to $\W[y]$ is $\F[2]$, as $\{ P_{\W[1]} , P_{\F[2]} \} = 0$. At this point, we can determine the last two glues,
\begin{gather}
    P_{\W[y]} \, P_{\D[1]{1}} \, P_{\F[2]} = - \Gamma^{0y12} \, i \sigma_2 = - P_{\D[y12]{3}} \\
    P_{\W[y]} \, P_{\W[1]} \, P_{\D[134]{3}} = \Gamma^{0y34} \, i \sigma_2 = P_{\D[y34]{3}} \, .
\end{gather}
We discard the sign in front of $P_{\D[y12]{3}}$, as the orientation is fixed by the coefficients. The third pair of glue is, therefore, a pair of $\D{3}$-branes. The resulting configuration can be seen on the right-hand side of figure \ref{fig:D1-D5-P-Triangle}. Analogously, the third pair of glues for the first case is $\NS[y1234]$-$\F[1]$, and its whole configuration is displayed on the left-hand side of figure \ref{fig:D1-D5-P-Triangle}.

\begin{figure}[ht]
    \centering
    \includegraphics[width = 0.4 \textwidth]{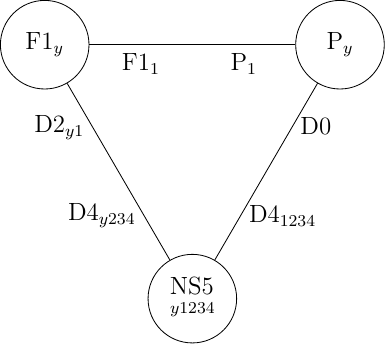}
    \caption{Graphical representation of the type IIA super-maze.}
    \label{fig:super-maze}
\end{figure}

The super-maze is another important example of a 1/8-BPS system with 16 local supercharges. It is dual to both configurations. Performing an S-duality on both configurations, followed by a T-duality along $x^1$ for the left and along $x^2$ for the right configuration in figure \ref{fig:D1-D5-P-Triangle}, we obtain the Type-II super-maze displayed in figure \ref{fig:super-maze}. It was first derived in \cite{Bena:2022wpl} to track the degrees of freedom of a $\NS$-$\F$-$\W$ black hole. Its entropy in the $g_S N \to 0$ limit is generated by fractionation of the $N_1$ fundamental strings into $N_1 N_5$ little strings inside the worldvolume of the $\NS$ brane. This is best understood in its M-theory uplift, where $N_1 N_5$ little $\M{2}$-strips are stretched between the $\M{5}$ branes. Turning on the interaction between $\M{2}$ and $\M{5}$ branes (or between $\F$-strings and $\NS$-branes in the Type-IIA frame), the $\M{2}$ little strips begin to pull on the $\M{5}$ branes, creating a Callan-Maldacena-like furrow. A great illustration of the fractionation and back-reaction of the branes can be found in figures 1 and 3 in \cite{Bena:2022wpl}.

\section{Symmetries for main branes and glues}\label{sec:3ChargeSymmetries}

\begin{figure}[htb]
    \centering
    \begin{subfigure}{0.45\textwidth}
        \includegraphics{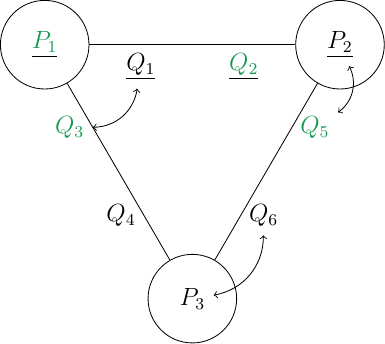}
    \end{subfigure}
    \begin{subfigure}{0.45\textwidth}
        \includegraphics{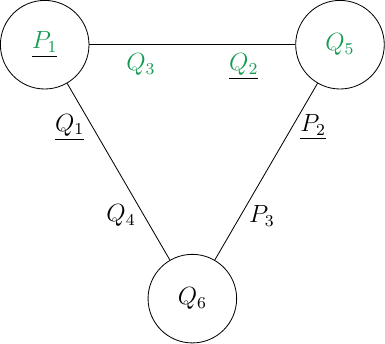}
    \end{subfigure}
    \caption{The left figure shows the three-charge configuration before the transformation, the right figure after. The arrows indicate which branes switch roles. 
    The involutions in $\mu_{15} = \color{ForestGreen}{P_1 Q_2 Q_3 Q_5}$ are highlighted in green and the involutions in $\eta_{12} = \underline{P_1 Q_1 Q_2 P_2}$ are underlined. This highlights the exchange of roles of $\eta_{12}$ and $\mu_{15}$.}
    \label{fig:PermutationSymmetryFor3Charges}
\end{figure}

As for the 2-charge case, there is a symmetry for the existence of a local supersymmetry enhancement for 3-charge configuration. Given a system as discussed above, we aim to interchange the glues $Q_5$ and $Q_6$ with the main branes $P_2$ and $P_3$. This preserves the enhancement of the subsystem given by $\eta_{23}$, but it breaks it for $\mu_{15}$ and $\mu_{16}$. We denote the system after the exchange with a prime, i.e. $P_2 ' = Q_5$ , $Q_6 ' = P_3$, and so forth. If we further exchange the glues $Q_1$ and $Q_3$, we have $\eta_{12} ' = P_1 Q_3 Q_2 Q_5 = \mu_{15}$ and $\mu_{15} ' = P_1 P_2 Q_1 Q_2 = \eta_{12}$, and analogously for $\eta_{13} '$ and $\mu_{16}'$ (see figure \ref{fig:PermutationSymmetryFor3Charges}). Thus, the new system has signs $\eta_{12} '$, $\eta_{13} '$, $\mu_{15} '$ and $\mu_{16} '$, along with the correct (anti-) commutation relations. Therefore, it also admits local supersymmetry enhancement.

\begin{figure}[htb]
    \centering
    \begin{subfigure}{0.4\textwidth}
        \includegraphics{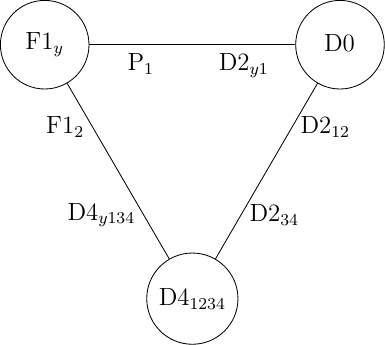}
        \caption{Can be interpreted as a supertube smoothly ending on a $\D{4}$ brane, similar to a Callan-Maldacena spike.}
        \label{fig:F1-D0-D4-1}
    \end{subfigure}
    \hfill
    \begin{subfigure}{0.4\textwidth}
        \includegraphics{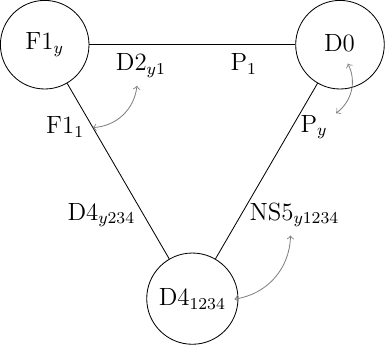}
        \caption{Applying the symmetry with $\F[y]$ fixed (\textcolor{gray}{gray} arrows) gives the super-maze in figure \ref{fig:super-maze}.}
        \label{fig:F1-D0-D4-2}
    \end{subfigure}
    
    \caption{Possible local supersymmetry enhancement of $\D{0}$-$\F$-$\D{4}$ with $\D{2}$-$\D{2}$ glues on the left and $\NS$-$\W$ glues on the right.}
    \label{fig:F1-D0-D4-possibilities}
\end{figure}

An illustration of this can be found when considering a system with the following main charges: a fundamental string $\F[y]$ perpendicular to a $\D[r \phi 34]{4}$ brane on a torus $T^4_{r \phi 34}$, together with $\D{0}$-charges. There are two different sets of glues possible. The first one, displayed in figure \ref{fig:F1-D0-D4-1}, has the simple interpretation of supertubes pulling on a $\D{4}$-brane. The $\F$ and $\D{0}$ form together with the glues $\W[\phi]$ and $\D[y\phi]{1}$ a supertube, where $\phi$ is the azimuth coordinate in the torus. The $\F$-string furthermore pulls on the $\D{4}$-brane forming a Callan-Maldacena spike but for a supertube. Besides the usual $\F[r]$ and $\D[y \phi 34]{4}$ dipole charges that allow a smooth transition between the string and the D-brane, we have $\D[r \phi]{2}$ and $\D[34]{2}$ glues as part of the $\D{0}$-$\D{4}$ subsystem. The first one enables the $\D[y \phi]{2}$-tube to curve into the $\D{4}$-brane, while the second one makes sure to retain 16 local supercharges. Altogether, we have a coalescence of a supertube with a Callan-Maldacena spike, while retaining preserving 16 local supercharges. 

The second one cannot be interpreted in this way. It uses the same direction to smooth the transition between the F-string and $\D{4}$-brane and to expand into a supertube. As a result it uses $\NS[y1234]$-$\W[y]$ glues between the $\D{0}$ and $\D{4}$ brane. Its interpretation is not apparent in the Type IIA frame. Instead, the authors in \cite{Eckardt:2023nmn} consider its M-theory uplift and discover a super-maze in the $(x^{11},x^1)$ plane, where $x^{11}$ is the M-theory direction. Returning to the topic on hand, this configuration can be found without following the procedure presented in the previous section. Instead, we can use the symmetry discussed above on the super-maze in figure \ref{fig:super-maze}. We interchange the main branes $\NS[y1234]$ and $\W[y]$ on the right side with their glues $\D[1234]{4}$ and $\D{0}$, as well as the glues $\F[1]$ and $\D[y1]{2}$. This results in the $\F$-$\D{0}$-$\D{4}$ configuration with $\NS$-$\W$ glues displayed in figure \ref{fig:F1-D0-D4-2}. Another simple way to generate both configurations is a T-duality along $y$ of the $\D{1}$-$\D{5}$-$\W$ systems discussed in the previous section (see figure \ref{fig:D1-D5-P-Triangle}).

\section{A tetrahedral ansatz for a four-charge system}
\label{sec:tetrahedralAnsatz-4Charges}

\begin{figure}
    \centering
    \includegraphics[width = 0.4 \textwidth]{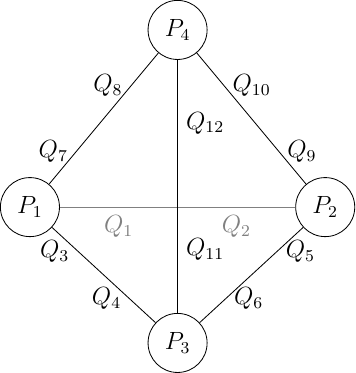}
    \caption{Tetrahedral ansatz for a 4-charge black hole. Each one of the four facets of the tetrad forms an enhanced 3-charge system by satisfying the conditions for the triangle ansatz.}
    \label{fig:tetrahedonansatz}
\end{figure}

Based on the logic of the previous sections, the most natural way to enhance the supersymmetry of a four-charge system is by a tetrahedral structure, where each facet represents the local supersymmetry enhancement of a three-charge system (see figure \ref{fig:tetrahedonansatz}). We show that this is \textbf{not} a correct ansatz.

Let us build up the tetrahedron facet by facet. We start with the bottom and back side and assume 3-charge local supersymmetry enhancements.
This corresponds to enhancing the $P_1$-$P_2$-$P_3$ and $P_1$-$P_2$-$P_4$ subsystems with the signs
\begin{align}
    \mu_{1,5} &= P_1 Q_2 Q_3 Q_5 \\
    \mu_{1,6} &= P_1 Q_1 Q_4 Q_6 \\
    \mu_{1,9} &= P_1 Q_2 Q_7 Q_9 \\
    \mu_{1,10} &= P_1 Q_1 Q_8 Q_{10} \, .
\end{align}

The third, left, facet is already ambiguous in the way we can define the $\mu$s. There are two ways, denoted by $\mu$ and $\mu'$.

\begin{align}
    \mu_{1,11} &= P_1 Q_4 Q_7 Q_{11} \\
    \mu_{1,12} &= P_1 Q_3 Q_8 Q_{12} \\
    \mu'_{1,11} &= P_1 Q_3 Q_7 Q_{11} \\
    \mu'_{1,12} &= P_1 Q_4 Q_8 Q_{12} 
\end{align}

The enhancement of the last facet should then follow from the others. Trying out all possibilities, there is only one combination that works (up to exchanging $Q_{11}$ and $Q_{12}$) using the unprimed $\mu$s.
\begin{align}
    \mu_{1,5} \, \mu_{1,9} \, \mu_{1,11} \eta_{13} &= 
    - P_3 Q_5 Q_{11} Q_9 =: - \mu_{3,9}
\end{align}
All the other combinations fail because a product of $\eta$s and $\mu$s would result in a product of two involutions. This automatically demands that these two are equivalent up to a sign, altering the structure and breaking the symmetry of the ansatz.

At this point, we would have to cancel the new anticommutators between glues on opposite edges with each other and may have to introduce new equations for this. While this is possible\footnote{The only new sign that has to be introduced to cancel the anticommutators is $\lambda := P_1 P_2 P_3 P_4 = \pm 1$.}, it is not relevant as the equations that follow from the 3-charge local supersymmetry enhancement of the facets already forbid a solution. 
All equations but the BPS condition remain the same for the facets as for the 3-charge configuration. Therefore, with the analogous ansatz, $\alpha_1 = a^2$, $\alpha_2 = b^2$, $\alpha_3 = bc^2$ and $\alpha_4 = d^2$, we can directly use the solution \eqref{eq:Solution-TriangleAnsatz} for the 3-charge case for the coefficients $\beta_i$:
\begin{align}
    \beta_1 &= \kappa_1 \, a b \\
    \beta_3 &= \kappa_3 \, a c \\
    \beta_5 &= \kappa_5 \, b c \\
    \beta_7 &= \kappa_7 \, a d \\
    \beta_9 &= \kappa_9 \, b d \\
    \beta_{11} &= \kappa_{11} \, c d
\end{align}
with the constraints
\begin{align*}
    \kappa_1 \kappa_3 \kappa_5 &= - \mu_{1,5} \\
    \kappa_1 \kappa_7 \kappa_9 &= - \mu_{1,9} \\
    \kappa_3 \kappa_7 \kappa_{11} &= \mu_{1,11} \eta_{13} \\
    \kappa_5 \kappa_9 \kappa_{11} &= - \mu_{3,9}
\end{align*}
If we multiply all of these constraints, we find
\begin{align}
    (\kappa_1 \kappa_3 \kappa_5 \kappa_7 \kappa_9 \kappa_{11})^2 = - \mu_{1,5} \mu_{1,9} \mu_{1,11} \eta_{13} \mu_{3,9}
\end{align}
Expanding the right-hand side, we find that all projectors cancel, giving us $-1$. This is a contradiction, as the left-hand side is positive. The equations do not have a solution.

  \chapter{The hanger ansatz}\label{chap:HangerAnsatz}

Beside the triangle ansatz, there is another ansatz that has 16 supercharges locally, the \textit{hanger ansatz}. 
It realizes the idea of a 3-charge enhancement by connecting all main branes by one glue each. Graphically this represents a 3-pointed star and is discussed in section \ref{sec:3pointedStar}. As this ansatz alone has no solution, we add another pair of glues, resulting in the hanger ansatz in section \ref{sec:HangerAnsatz}. Two examples of hanger configuration are given in section \ref{sec:ExamplesHanger}. We conclude by generalizing it to four main charges and again give two examples.

\section{The 3-pointed star}\label{sec:3pointedStar}

\begin{figure}
    \centering
    \includegraphics[width = 0.4 \textwidth]{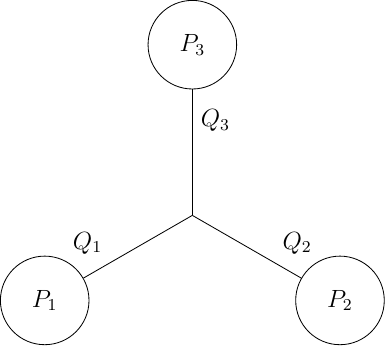}
    \caption{Graphical representation of the 3-pointed star. Any two glues form a 2-charge local supersymmetry enhancement between their main branes.}
    \label{fig:3pointedStarAnsatz}
\end{figure}

The 3-pointed star is a subsystem of the hanger ansatz and the main point in which they differ. Its equations do not have a solution, but we modify them in the next section to give rise to the hanger ansatz.
Instead of a pair of glues for each pair of main branes, we associate only one glue to each main brane. 
\begin{align}
    f_i &:= \alpha_i + \beta_i Q_i P_i \, .
\end{align}
The idea is that the anticommutator of two main branes cancels with the anticommutator of its two glues. That way each glue combines with any of the other two glues to enhance the local supersymmetry of their main brane's subsystem, see figure \ref{fig:3pointedStarAnsatz}. For the triangle ansatz, further anticommutators between glues on different sides of the triangle forced us to introduce a more complicated structure. This is not the case for the 3-pointed star, as it uses only half of the glues. All pairs of branes are contained in one of the three subsystems given by the signs,
\begin{align}
    \eta_{12} &:= P_1 Q_1 Q_2 P_2 \\
    \eta_{13} &:= P_1 Q_1 Q_3 P_3 \\
    \eta_{23} &:= P_2 Q_2 Q_3 P_3 = \eta_{12} \, \eta_{13} \, .
\end{align}

The (anti-) commutation relations are the usual ones for the constituents of the signs. Glues commute with glues, and main branes anticommute with glues. As the main branes commute to preserve some supersymmetry globally, these relations follow from only one anticommutator, e.g.
\begin{align}
    \{ P_1, Q_1 \} = 0 \, .
\end{align}

Inserting the functions $f_i$ into the global supersymmetry condition in equation \eqref{eq:globalsupersymmetrycondition}, a third glue charge is added to equation \eqref{eq:betaRelation-2Charge},
\begin{align}
    0 &= \beta_1 + \eta_{12} \beta_2 + \eta_{13} \beta_{3} \, . \label{eq:betaRelations-3PointedStar}
\end{align}
The BPS-equation, as well as the three equations to cancel the anticommutators are the usual BPS-equation for three main charges,
\begin{align}
    1 &= \alpha_1 + \alpha_2 + \alpha_3 \label{eq:BPSCondition-3PointedStar}\\
    \alpha_1 \alpha_2 &= - \eta_{12} \beta_1 \beta_2 \label{eq:FirstCancellationEquation-3PointedStar}\\
    \alpha_1 \alpha_3 &= - \eta_{13} \beta_1 \beta_3 \\
    \alpha_2 \alpha_3 &= - \eta_{23} \beta_2 \beta_3 \, .
\end{align}
As was the case with the triangle ansatz, the first projector condition \eqref{eq:firstProjectorCondition} is redundant,
\begin{align*}
    1 &= 1^2 - 0^2 = \left( \sum_i \alpha_i \right)^2 - \left( \sum_j \beta_j Q_j P_j \right)^2 \\
    &= \sum_i \alpha_i^2 + 2 \, \alpha_1 \alpha_2 + 2 \, \alpha_1 \alpha_3 + 2 \, \alpha_2 \alpha_3 \\
    & \quad + \sum_j \beta_j^2 + 2 \, \eta_{12} \, \beta_1 \beta_2 + 2 \, \eta_{13} \, \beta_1 \beta_3 + 2 \, \eta_{23} \, \beta_2 \beta_3 \\
    &= \sum_i \alpha_i^2 + \sum_j \beta_j^2 \, .
\end{align*}
Beside the BPS-equation, we derived four equations for three glue charge densities. This system of equations is overdetermined and does \textbf{not} have a solution. The difference to the tetrad ansatz is that the equations to cancel the anticommutators are not altered due to the extra structure from the $\mu$s. This allows us to add another term to them in the next section.

\section{The hanger ansatz}\label{sec:HangerAnsatz}

\begin{figure}[ht]
    \centering
    \includegraphics[width = 0.5 \textwidth]{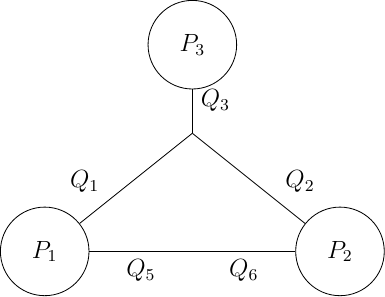}
    \caption{A graphical representation of the hanger ansatz. It consists of a 3-pointed star connecting all main branes, and an extra pair of glues $Q_5$-$Q_6$ at the bottom.}
    \label{fig:HangerAnsatz}
\end{figure}

The goal is to modify the equations of the 3-pointed star ansatz. Equation \eqref{eq:BPSCondition-3PointedStar} is the BPS-equation for three main charges and equation \eqref{eq:betaRelations-3PointedStar} the typical equation of this ansatz. It does not make sense to modify them. The remaining equations come from the cancellation of anticommutators in the second projector condition \eqref{eq:secondProjectorCondition}. The idea is to add another anticommutator $\{ Q_5 , Q_6 \}$ that is proportional to one of the other anticommutators. This amounts to adding a pair of glues $Q_5$ and $Q_6$ between two main branes, giving the ansatz the shape of a hanger, see figure \ref{fig:HangerAnsatz}.

The ansatz for the functions $f_i (x)$ becomes
\begin{subequations}
\begin{align}
    f_1 &:= \alpha_1 + \beta_1 Q_1 P_1 + \beta_5 Q_5 P_1 \\
    f_2 &:= \alpha_2 + \beta_2 Q_2 P_2 + \beta_6 Q_6 P_2 \\
    f_3 &:= \alpha_3 + \beta_3 Q_3 P_3 \, .
\end{align}
\end{subequations}
Beside the usual relation for a pair of glues (see equation \eqref{eq:betaRelation-2Charge}), the difference to the 3-pointed star is the cancellation of the anticommutator $\{ P_1 , P_2 \}$. As the anticommutator of the extra pair of glues is proportional to it, we have three terms that cancel each other,
\begin{align}
    \alpha_1 \alpha_2 \{ P_1 , P_2 \} + \beta_1 \beta_2 \{ Q_1 , Q_2 \} + \beta_5 \beta_6 \{ Q_5 , Q_6 \} = 0 \, .
\end{align}
This alters equation \eqref{eq:FirstCancellationEquation-3PointedStar} by adding another positive term. The equations for the hanger ansatz are, therefore, 
\begin{align}
    1 &= \alpha_1 + \alpha_2 + \alpha_3 \\
    0 &= \beta_1 + \eta_{12} \beta_2 + \eta_{13} \beta_3 \\
    0 &= \beta_5 + \eta'_{12} \beta_6 \\
    0 &= \alpha_1 \alpha_2 + \eta_{12} \beta_1 \beta_2 + \eta'_{12} \beta_5 \beta_6 \\
    0 &= \alpha_1 \alpha_3 + \eta_{13} \beta_1 \beta_3 \\
    0 &= \alpha_2 \alpha_3 + \eta_{23} \beta_2 \beta_3 \, ,
\end{align}
where the sign $\eta_{12}' := P_1 Q_5 Q_6 P_2$ is the usual sign for a 2-charge local supersymmetry enhancement.

To solve these equations, we make the usual ansatz, $\alpha_1 = a^2$, $\alpha_2 = b^2$ and $\alpha_3 = c^3$, and choose any parametrization of the 2-sphere for $a$, $b$ and $c$. 
This gives the same two parameters we can vary in space as for the triangle ansatz.
The coefficients for the glues are given by
\begin{subequations}\label{eq:solution-hanger}
\begin{align}
    \beta_1 &= - \kappa_3 \eta_{13} \, \frac{a^2 c}{\sqrt{a^2+b^2}} \\
    \beta_2 &= - \kappa_3 \eta_{23} \, \frac{b^2 c}{\sqrt{a^2+b^2}} \\
    \beta_3 &= \kappa_3 \, \frac{(a^2 + b^2) \, c}{\sqrt{a^2 + b^2}} \\
    \beta_5 &= \kappa_5 \frac{a b}{\sqrt{a^2 + b^2}} \\
    \beta_6 &= - \eta_{12}' \, \beta_5 \, ,
\end{align}
\end{subequations}
where $\kappa_3 = \pm 1$ and $\kappa_5 = \pm 1$ are signs, giving some freedom of choice for the orientation of the glues.

Up until now, we have ignored the main problem of freely adding glues to the system. The anticommutators with other glues and main branes also appear in the second projector condition\eqref{eq:secondProjectorCondition}. In this case, $Q_5$ and $Q_6$ have to anticommute with all other glues and $P_3$. This imposes a heavy constraint on $Q_5$ and $Q_6$, and often requires external degrees of freedoms.

\section{Examples for hanger configurations}\label{sec:ExamplesHanger}

\begin{figure}[htb]
    \centering
    \includegraphics[width = 0.5 \textwidth]{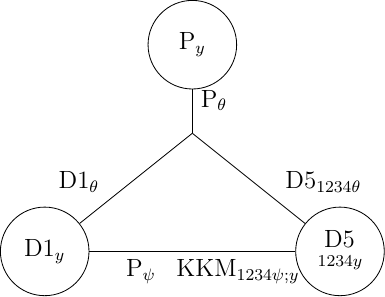}
    \caption{Diagram displaying the brane charges of the superstratum in \cite{Bena:2011uw}}
    \label{fig:D1-D5-P-Hanger}
\end{figure}

The most prominent example of a hanger configuration is the superstratum discussed in \cite{Bena:2011uw}. Its main charges are a $\D[y]{1}$ string with momentum $\W[y]$ on a circle $S^1_y$, and a $\D[y1234]{5}$ brane wrapping the same circle and a $T^4_{1234}$. The structure of the three pointed star is achieved by using the same charges as the main branes for the glues, but with the direction $z$ replaced by an external direction $\theta$. For the additional pair of glues, the authors of \cite{Bena:2011uw} used a $\KKM[1234\psi;z]$ along the $T^4$ and an external direction $\psi$ together with a momentum along $\psi$. The special direction of the $\KKM$ is $y$. With two external directions $\theta$ and $\psi$, this gives a two-dimensional sheet in the transverse space $\mathbb{R}^4$. The whole configuration is displayed in figure \ref{fig:D1-D5-P-Hanger}.

\begin{figure}[htb]
    \centering
    \includegraphics[width = 0.5 \textwidth]{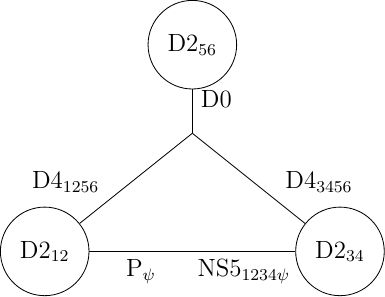}
    \caption{$\D[12]{2}$-$\D[34]{2}$-$\D[56]{2}$ hanger configuration with one external direction $\psi$.}
    \label{fig:D2-D2-D2-Hanger}
\end{figure}

Other examples can be constructed by dualities, for example the $\D[12]{2}$-$\D[34]{2}$-$\D[56]{2}$ system displayed in figure \ref{fig:D2-D2-D2-Hanger}. It uses only one external direction $\psi$, while it is still determined by functions of two variables, like the superstratum.

\section{The 4-charge hanger ansatz}\label{sec:4chargeHanger}

\begin{figure}[htb]
    \centering
    \resizebox{0.5 \textwidth}{!}{
    }
    \includegraphics[width = 0.5 \textwidth]{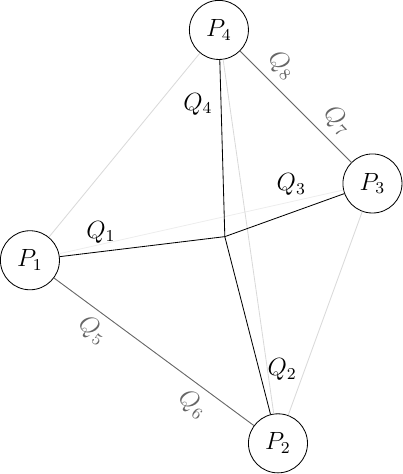}
    \caption{4-charge hanger configuration. It consists of a four-pointed star and two additional pairs of glues in dark gray. Removing any main brane $P_i$ together with its glues gives a 3-charge hanger configuration.}
    \label{fig:4-Charge-Hanger}
\end{figure}

Contrary to the triangle ansatz, the hanger ansatz can be generalized to four charges. The starting point is a 4-pointed star. This gives us for each main charge $P_i$ a glue $Q_i$, such that we can define the signs
\begin{subequations}
\begin{align}
    \eta_{12} := P_1 Q_1 Q_2 P_2 = \pm 1 \\
    \eta_{13} := P_1 Q_1 Q_3 P_3 = \pm 1 \\
    \eta_{14} := P_1 Q_1 Q_4 P_4 = \pm 1 \, .
\end{align}
\end{subequations}
Then, again, its equations have no solution. They have to be altered by adding further pairs of glues for certain main branes. To determine them, we consider parts of the system. For the subsystem $P_1$-$P_2$-$P_3$, introducing the pair $Q_5$-$Q_6$ between $P_1$ and $P_2$ makes this a 3-charge hanger. For $P_2$-$P_3$-$P_4$, we need another pair $Q_7$-$Q_8$ between $P_3$ and $P_4$. Thus, two more signs commence
\begin{subequations}
\begin{align}
    \eta_{12}' := P_1 Q_5 Q_6 P_2 \\
    \eta_{34}' := P_3 Q_7 Q_8 P_4 \, .
\end{align}
\end{subequations}
These two pairs of glues ensure, when disregarding any main charge (together with its glues), the remaining subsystem is a 3-charge hanger configuration, see figure \ref{fig:4-Charge-Hanger}. Conversely, as for the 3-charge case, when adding further glues, they have to anticommute with every other brane. Often, this can only be done by utilizing extra dimensions, as is the case for the $\D{1}$-$\D{5}$-$\W$-$\KKM$ system below.

We skip the detailed computations and only present the resulting equations governing the 4-charge hanger.
\begin{subequations}
\begin{align}
    1 &= \alpha_1 + \alpha_2 + \alpha_3 + \alpha_4 \label{eq:4-Charge-Hanger-BPS}\\
    0 &= \beta_1 Q_1 P_1 + \beta_2 Q_2 P_2 + \beta_3 Q_3 P_3 + \beta_4 Q_4 P_4 \label{eq:4-Charge-Hanger-betaRelations}\\
    0 &= \alpha_1 \alpha_2 + \eta_{12} \beta_1 \beta_2 - \beta_5^2 \label{eq:4-Charge-Hanger-alphaBetaRelations-1}\\
    0 &= \alpha_1 \alpha_3 + \eta_{13} \beta_1 \beta_3 \\
    0 &= \alpha_1 \alpha_4 + \eta_{14} \beta_1 \beta_4 \\
    0 &= \alpha_2 \alpha_3 + \eta_{23} \beta_2 \beta_3 \\
    0 &= \alpha_2 \alpha_4 + \eta_{24} \beta_2 \beta_4 \\
    0 &= \alpha_3 \alpha_4 + \eta_{34} \beta_3 \beta_4 - \beta_7^2 \label{eq:4-Charge-Hanger-alphaBetaRelations-6}\\
    0 &= \beta_5 + \eta_{12}' \beta_6 \\
    0 &= \beta_7 + \eta_{34}' \beta_8 
\end{align}
\end{subequations}
The first equation, \eqref{eq:4-Charge-Hanger-BPS}, is the BPS-equation for four charges, the second one, \eqref{eq:betaRelations-3PointedStar} the typical equation for a 4-pointed star. Notice the extra terms in equations $\eqref{eq:4-Charge-Hanger-alphaBetaRelations-1}$ and \eqref{eq:4-Charge-Hanger-alphaBetaRelations-6} from the additional two pairs of glues.

A solution can be found by the same trick as for 3-charge configurations. With the substitution $\alpha_1 = a^2$, $\alpha_2 = b^2$, $\alpha_3 = c^3$ and $\alpha_4 = d^2$, the first equation becomes the equation for a 3-sphere. Choosing any parametrization, we find three degrees of freedom to vary in space. The coefficients for the glues are
\begin{subequations}
\begin{align}
    \beta_1 &= \kappa_1 \, a^2 \sqrt{\frac{c^2 + d^2}{a^2 + b^2}} \\
    \beta_2 &= \kappa_1 \eta_{12} \, b^2 \sqrt{\frac{c^2+d^2}{a^2+b^2}} \\
    \beta_3 &= - \kappa_1 \eta_{13} \, c^2 \sqrt{\frac{a^2+b^2}{c^2+d^2}} \\
    \beta_4 &= - \kappa_1 \eta_{14} \, d^2 \sqrt{\frac{a^2+b^2}{c^2+d^2}} \\
    \beta_5 &= \kappa_5 \sqrt{\frac{a^2 b^2}{a^2 + b^2}} \\
    \beta_6 &= - \eta_{12}' \, \beta_5 \\
    \beta_7 &= \kappa_7 \sqrt{\frac{c^2 d^2}{c^2 + d^2}} \\
    \beta_8 &= - \eta_{34}' \, \beta_7 \, ,
\end{align}
\end{subequations}
with signs $\kappa_1 , \kappa_5 , \kappa_7 = \pm 1$ that describe some freedom in choice for the orientation of glues.

\section{Two instances of 4-charge hangers}\label{sec:Examples4chargeHanger}

\begin{figure}[htb]
    \centering
    \includegraphics[width = 0.5 \textwidth]{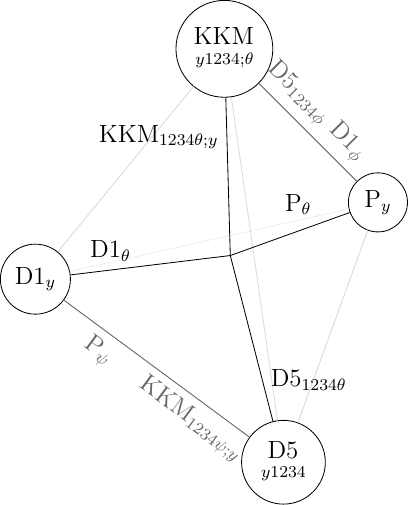}
    \caption{4-charge configuration adding a $\KKM$ charge to the superstratum in \cite{Bena:2011uw}. The $\D[\phi]{1}$-$\D[1234\phi]{5}$ glues can be exchanged for $\W[\phi]$-$\KKM[1234\phi]$ glues, and vice versa.}
    \label{fig:D1-D5-P-KKM}
\end{figure}

The question arises whether the superstratum found in \cite{Bena:2011uw} can be generalised to four charges, and indeed, this is the case. A fourth main charge $\KKM[y1234;\theta]$ along the same directions as the $\D{5}$ can be added and glued to the system with $\KKM[1234\theta;y]$, see figure \ref{fig:D1-D5-P-KKM}.\footnote{While the special direction of the $\KKM$s cannot be deduced by supersymmetry arguments, we have chosen them such that this configuration is dual to the $\D{2}$-$\D{2}$-$\D{2}$-$\D{6}$ configuration below.} For the two extra glues, we have two choices each, $\W[\psi_1]$-$\KKM[1234\psi_1;y]$ or $\D[\psi_2]{1}$-$\D[1234\psi_2]{5}$, where $\psi_1$ and $\psi_2$ are directions in the transverse space. This direction has to be different for the two pairs of glues to ensure that they anticommute, and different to the common direction $\theta$ of the other glues. In total, we require three of the four transverse direction, $\theta$, $\psi$ and $\phi$. The configuration forms a three-dimensional hypersurface in the non-compact directions, $\mathbb{R}^4$. 

Despite adding a fourth charge, this configuration is still 1/8-BPS like its 3-charge counterpart. As $P_4 = P_1 P_2 P_3$, it follows that the fourth charge does not break any more supersymmetry,
\begin{equation}
    P_4 \, \epsilon = P_1 P_2 P_3 \, \epsilon = - \epsilon \, .
\end{equation}
Beware that this fixes the orientation of the $\KKM$. Flipping its orientation breaks all global supersymmetry.

\begin{figure}[tb]
    \centering
    \includegraphics[width = 0.5 \textwidth]{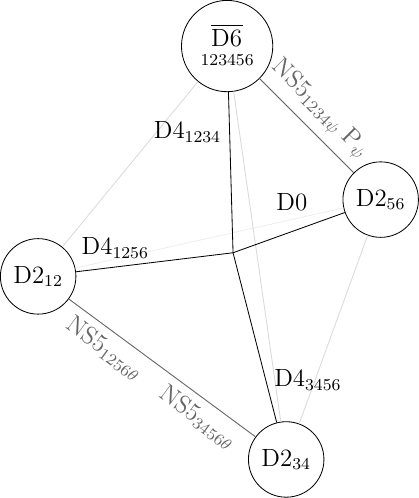}
    \caption{$\D{2}$-$\D{2}$-$\D{2}$-$\overline{\D{6}}$ hanger configuration. The $\NS[1256\theta]$-$\NS[3456\theta]$ glues can be exchanged by $\W[\theta]$-$\NS[1234\theta]$ glues, and vice versa.}
    \label{fig:D2-D2-D2-D6}
\end{figure}

Another configuration is the four dimensional $\D{2}$-$\D{2}$-$\D{2}$-$\D{6}$ black hole. It consists of three $\D{2}$ branes that extend along three different spheres $S^2_{12}$, $S^2_{34}$ and $S^2_{56}$, together with a $\D[123456]{6}$ anti-brane along all three spheres. They can be glued together by internal $\D{0}$ and $\D{4}$ branes, such that $P_i \, Q_i \propto \Gamma^{56} \, \sigma_3$, see figure \ref{fig:D2-D2-D2-D6}. As the main charges of the black hole span six dimensions, we have three remaining transverse direction, $\mathbb{R}^3$. The whole configuration is again a hypersurface of codimension $1$, i.e. a two-dimensional sheet. The extra pairs of glues require one transverse direction each. They can be chosen arbitrarily as either $\W[\psi_1]$-$\NS[1234\psi_1]$ or $\NS[1256\psi_2]$-$\NS[3456\psi_2]$ with transverse directions $\psi_1$ and $\psi_2$.
Beware the orientation of the $\D{6}$ anti-brane. As discussed, to preserve 1/8 of the supersymmetry globally, we have to choose $P_4 = P_1 P_2 P_3 = - \Gamma^{0123456} \, \sigma_1$.

  \chapter{Conclusion}\label{chap:Conclusion}

This thesis followed the idea of adding dipole charges to supersymmetric systems in order to preserve 16 supercharges locally. The result encompasses many configurations prominent in the literature, like the supertube, super-maze, or superstrata, but we also discovered new ways of enhancing those systems.

We first discussed the framework of local supersymmetry enhancement in chapter \ref{chap:LocalSupersymmetryEnhancement}. Applied to 1/4-BPS systems we derived general conditions on the glues, and succeedingly, applied them to find all pairs of glues for Type II string theory configurations in chapter \ref{chap:DualityMap}. This results in an enormous table of glues for configurations in different duality frames (see Appendix \ref{app:chap:dualityMapsWithGlues}). These are the building blocks of multi-charge systems with 16 local supercharges. 

In chapters \ref{chap:TriangleAnsatz} and \ref{chap:HangerAnsatz}, we combined three or four of these 1/4-BPS bound states to enhance three-charge systems corresponding to three-charge black holes. This leads to two different kinds of systems. The first follows from what we call the \textit{triangle} ansatz, where the main charges are connected by three pairs of glues. These glues are usually chosen to be inside the system, enhancing it without stretching out into the non-compact dimensions. The most prominent example is the super-maze \cite{Bena:2022wpl}, but there are many others. 
In particular, a back-reacted $\D{1}$-$\D{5}$-$\W$ configuration with exclusively internal glues has not been found before. 
We also tried to generalize the triangle ansatz to four charges, resulting in a tetrahedral structure, but the resulting equations admit no solution. 

In chapter \ref{chap:HangerAnsatz}, we present the other ansatz, that enhances the local supersymmetry of three-charge systems, the \textit{hanger} ansatz. It generalises the superstrata presented in \cite{Bena:2011uw} to different charges. It is inherently different from the triangle ansatz, not only in the number of different glues it uses, but also in the nature of the glues. 
The structure requires them to use directions transverse to the system, transforming it from a point-like structure in the transverse space into an extended, compact system. 
Furthermore, we discovered a set of glues to enhance four-charge systems by generalizing the hanger ansatz in the simplest way. The $\D{2}$-$\D{2}$-$\D{2}$-$\D{6}$ black hole, whose supergravity solution reduces to the extremal Reissner-Nordström black hole for equal charges, is given as one example of a 1/8-BPS system with four charges that can be enhanced to 16 local supercharges. This results in an extended, compact bound state with codimension 1.

In this thesis, we identified possible sets of glues to enhance a general 1/8-BPS system to preserve 16 supercharges locally. Together with the 'main branes', these dipole charges form pure bound states that we expect to describe the local picture of microstates of the corresponding black hole. The next step consists of confirming the existence of the bound states by constructing the supergravity solutions of these bound states, and identifying their duals in the corresponding conformal field theory. This has been achieved for some superstrata in \cite{Bena:2015bea}, but a construction for four-dimensional black holes and 4-charge systems like the $\D{2}$-$\D{2}$-$\D{2}$-$\D{6}$ has not been done yet. These two examples follow the structure of the hanger ansatz that generally break the rotational symmetry. Especially compelling is a possible supergravity solution of the super-maze, or any other configuration following the triangle ansatz. As these can use exclusively internal glues, they preserve the rotational symmetry.

This thesis also leaves another point unanswered. Is it possible to enhance a system with four main charges with exclusively internal glues, like the $\D{2}$-$\D{2}$-$\D{2}$-$\D{6}$ configuration? In section \ref{sec:4chargeHanger} we showed that the hanger ansatz generalizes to four charges naturally. The same failed for the triangle ansatz (see section \ref{sec:tetrahedralAnsatz-4Charges}). Although there was a way to naturally extend it to a tetrahedral structure, the resulting equations for the charge densities do not have a solution. We faced a similar problem when deriving the hanger ansatz. This was solved by modifying the equations with another pair of glues. For the tetrahedral structure, we did not find an apparent way to modify the equations to generate a solution.

As discussed before, the glues also give an idea on how the microstates of the black hole look like. This is excellently illustrated for the super-maze in \cite{Bena:2022wpl}. The glues directly lead to the interpretation of a smooth transition between the branes, leading to the $\M{5}$-$\M{2}$ furrow. Other examples include the supertube \cite{Mateos:2001qs} or the Callan-Maldacena spike \cite{Callan:1997kz}. However, this interpretation is not clear for all the 1/4-BPS building blocks discussed in section \ref{sec:ImportantConfigurations-2Charge}. Most notably are the two configurations below, as well as the analogous configurations when switching the 'main branes' with the glues. 
\begin{align*}
    [ \, \D{0} \perp \D[1234]{4} \; (0) \, ] \;
    &\rightarrow \;
    ( \, \D[12]{2} \perp \D[34]{2} \, ) \\
    [ \, \F[y] \perp \D[12345]{5} \; (0) \, ] \;
    &\rightarrow \;
    ( \, \D[1]{1} \perp \NS[y2345] \, ) 
\end{align*}
Particularly the 'inverse' of the former configuration is key to understanding the structure of the $\D{2}$-$\D{2}$-$\D{2}$-$\D{6}$ bound state presented in figure \ref{fig:D2-D2-D2-D6}.

In conclusion, we derived the dipole charges necessary for a supersymmetric system to preserve 16 supercharges locally. This resulted in two different glue structures, the first one using exclusively internal glues. 
These glues provide a first step towards discovering the structure of these bound states and constructing their supergravity solutions. 
We are eager to learn whether these microstates can account for the whole entropy of their corresponding black holes.


\vskip 20pt
\vskip 20pt
\noindent {\bf Acknowledgements:} 
I would like to thank my supervisors, Dr. Yixuan Li for the great discussions and immense help in this project, and PD Dr. Ralph Blumenhagen for the guidance and insightful lecture. I would also like to thank the Max Planck Institute for Physics for the funding of this thesis, as well as my family and friends for the incredible support over many years.
  \begin{appendix}

\chapter{Brane involutions}\label{app:chap:BraneInvolutions}

In this Appendix, we include the involutions of the brane types in Type II string theory considered in this thesis. The Gamma matrices denote the directions that the brane is extended along, while the Pauli matrix is specific for the type of brane. The branes are assumed to extend along the directions $x^1$, $x^2$, \ldots, $x^p$, where $p$ is the dimension of the brane.

\begin{align}
 P_{\W  }&=\Gamma^{01}&
 P_{\F }&=\Gamma^{01}\sigma_3\notag\\
 P_{\NS}^{\text{IIA}}&=\Gamma^{012345}&
 P_{\NS}^{\text{IIB}}&=\Gamma^{012345}\sigma_3\notag\\
 P_{\KKM}^{\text{IIA}}&=\Gamma^{012345}\sigma_3&
 P_{\KKM}^{\text{IIB}}&=\Gamma^{012345} =  \Gamma^{6789}\notag\\
 P_{\D{0}}&=\Gamma^0 i\sigma_2&
 P_{\D{1}}&=\Gamma^{01} \sigma_1\notag\\
 P_{\D{2}}&=\Gamma^{012} \sigma_1&
 P_{\D{3}}&=\Gamma^{0123} i\sigma_2\notag\\
 P_{\D{4}}&=\Gamma^{01234} i\sigma_2&
 P_{\D{5}}&=\Gamma^{012345} \sigma_1\notag\\
 P_{\D{6}}&=\Gamma^{0123456} \sigma_1&
 P_{\D{7}}&=\Gamma^{01234567} i\sigma_2\notag\\
 P_{\D{8}}&=\Gamma^{012345678} i\sigma_2&
 P_{\D{9}}&=\Gamma^{0123456789} \sigma_1\notag
 \label{variousProjectors}
\end{align}

\chapter{Local supersymmetry enhancements for the duality map}\label{app:chap:dualityMapsWithGlues}

\section{Notation}\label{sec:Notation}

In this section, we give a detailed description of the colouring and the notations used in the diagrams of the duality map.

Every node stands for a 1/4-BPS configurations. The colour indicates whether we are in a Type IIA (\textcolor{red}{red}) or Type IIB theory (\textcolor{yellow}{yellow}). An \textcolor{orange}{orange} node exists in both theories. The first part of the node describes the main charges of the configuration in the notation described in equation \eqref{eq:Notation-2Charge-Orthogonal} and \eqref{eq:Notation-2Charge-Parallel}. A star $*$ after the configuration indicate that the special direction of the KKM is along a direction of the second brane.
The second part lists all possible glues that enhance the configuration to 16 supercharges locally. They have to be understood modulo permutations of the coordinates. Orange nodes may have a third part. In that case, the second part lists all configurations in the Type IIA theory, and the third in the Type IIB theory.

The nodes are connected by T- and S-dualities. The index of the T-dualities denotes the direction of the duality. For the duality maps in chapter \ref{chap:DualityMap}, $T_\perp$ and $T_\parallel$ denotes whether the direction of the duality is along the directions of both branes, or not, $T_{\vdash}$ that the direction is parallel to the first brane and perpendicular to the second, and the other way around for $T_{\dashv}$. A star $*$ indicates that the duality is performed along the special direction of the KKM, and $\leftrightarrow$ that the two branes switch places.

\clearpage
\section{The building blocks along the Duality Map}
\label{sec:appendix_LSE_tables}

\subsection{P - Dp and F1 - Dq configurations}


\begin{figure}[ht!]
    \centering
    \includegraphics[width = \textwidth]{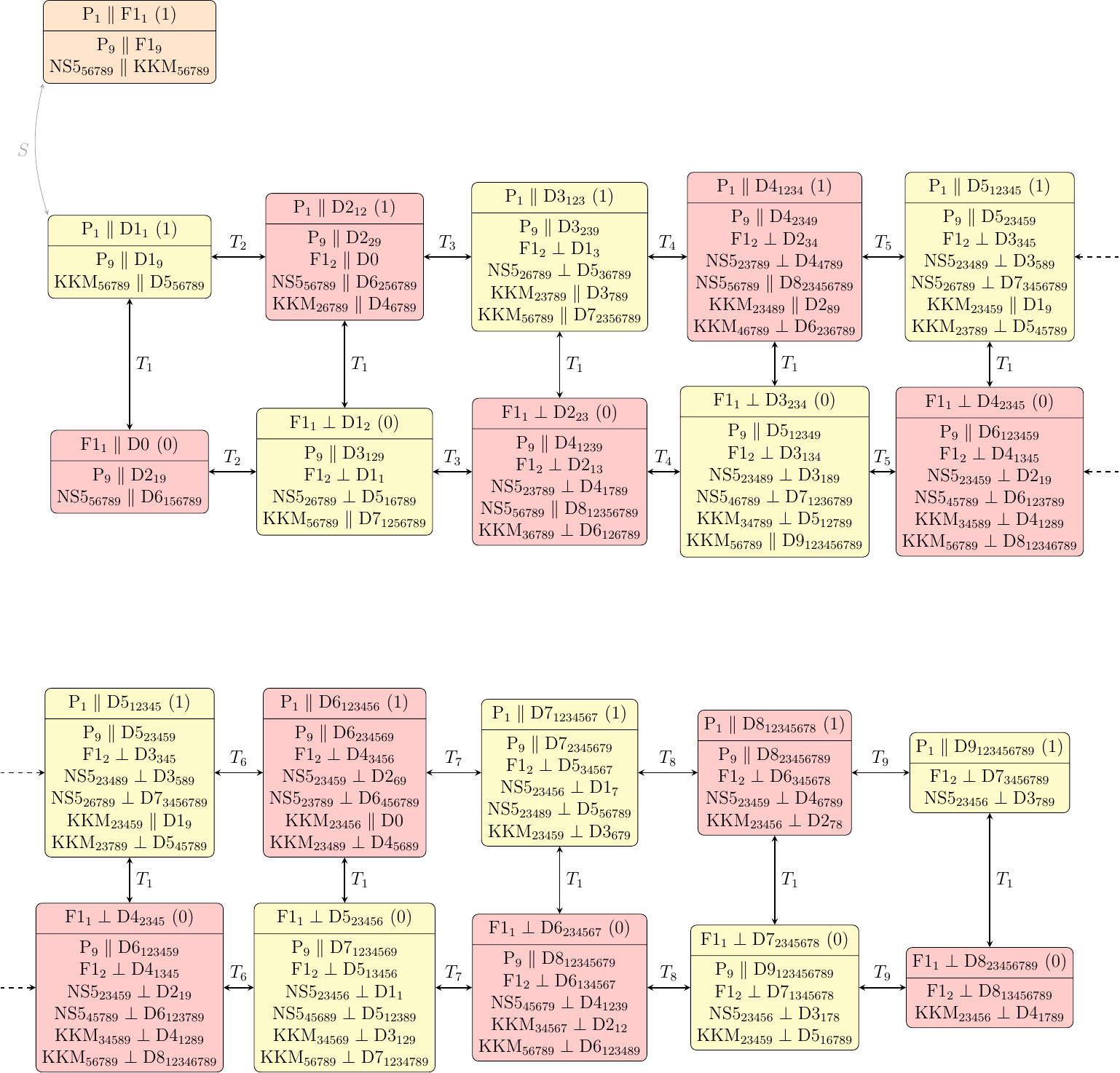}
    \caption{Possible local supersymmetry enhancements of $\W$-$\D{p}$ and $\F$-$\D{q}$ standard configurations. For details on colouring and notation, we refer to section \ref{sec:Notation}.}
    \label{fig:P-Dp-Configurations}
\end{figure}

\subsection{Dp - Dq configurations}

\begin{sidewaysfigure}[ht!]\footnotesize
    \centering
    \includegraphics[width = 0.9 \textwidth]{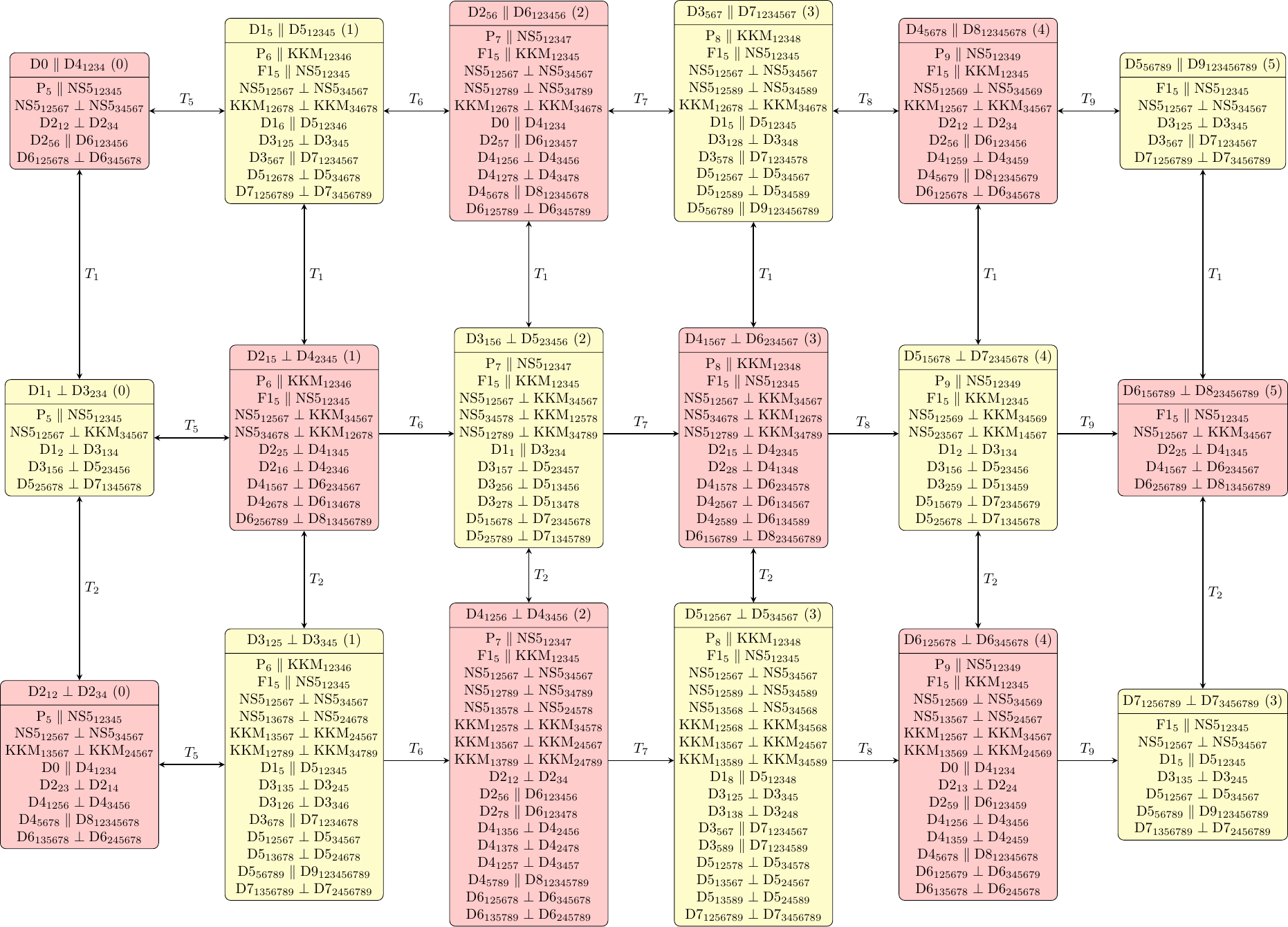}
    \caption{Possible local supersymmetry enhancements of $\D{p}$-$\D{q}$ standard configurations. For details on colouring and notation, we refer to section \ref{sec:Notation}.}
    \label{fig:Dp-Dq-Configurations}
\end{sidewaysfigure}

\clearpage
\subsection{Dp - NS5 and Dq - KKM configurations}


\begin{figure}[!ht]
    \centering
    \includegraphics[width = 0.7 \textwidth]{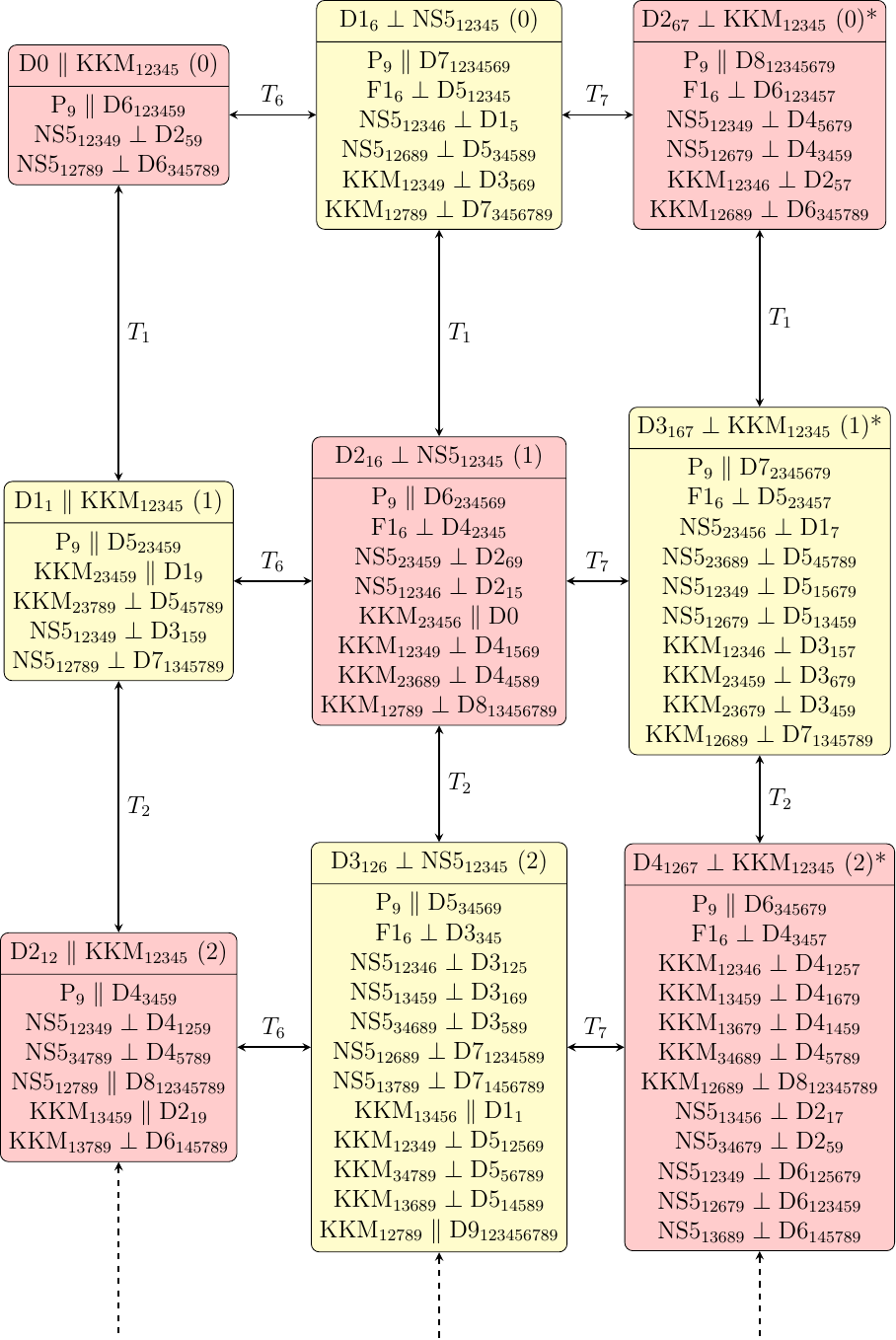}
    \caption{Possible local supersymmetry enhancements of $\D{p}$-$\NS$ and $\D{q}$-$\KKM$ standard configurations (first part). For details on colouring and notation, we refer to section \ref{sec:Notation}.}
    \label{fig:Dp-NS5-Configurations-1}
\end{figure}
\begin{figure}[!ht]
    \centering
    \includegraphics[width = 0.7 \textwidth]{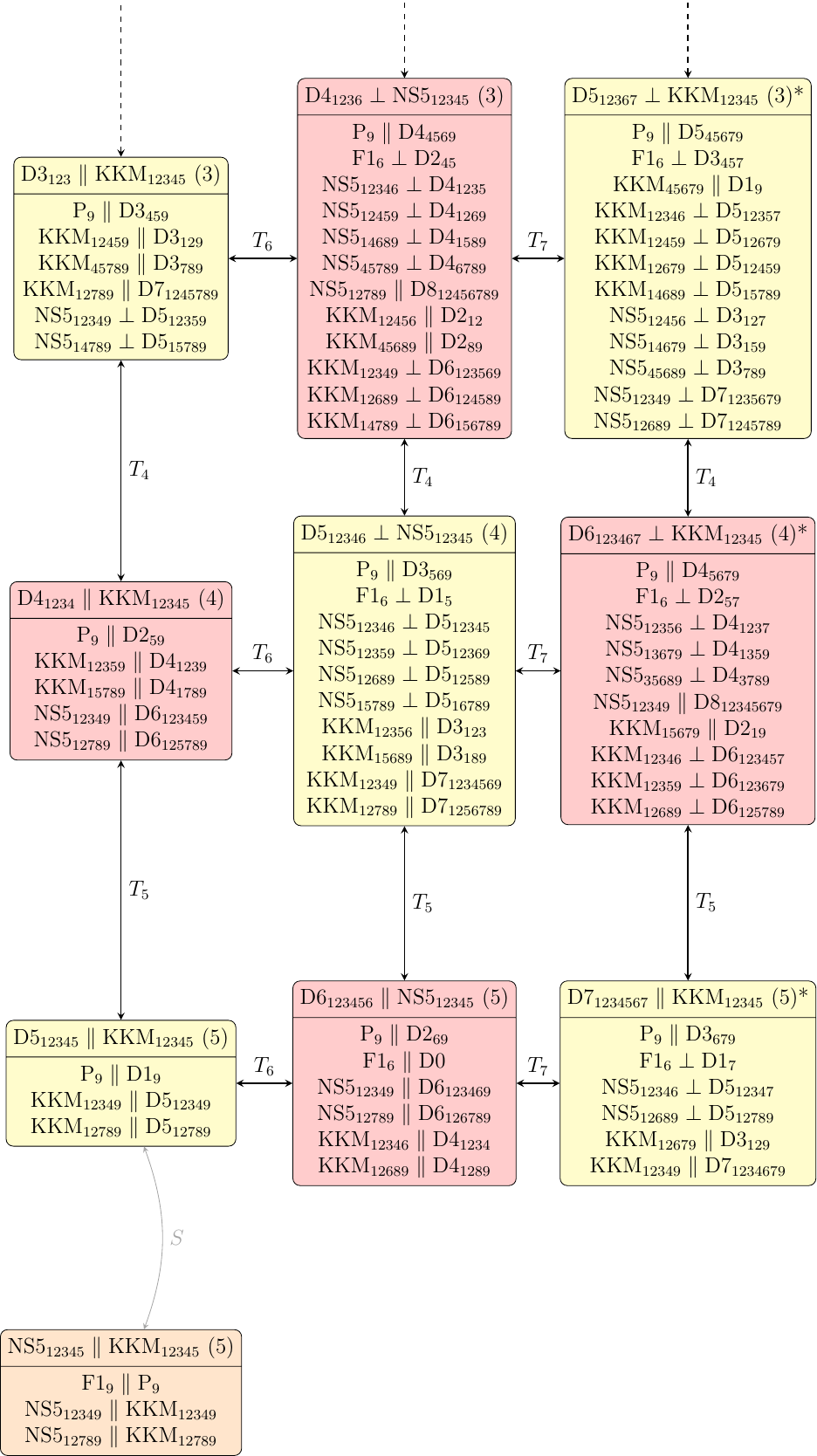}
    \caption{Possible local supersymmetry enhancements of $\D{p}$-$\NS$ and $\D{q}$-$\KKM$ standard configurations (second part). For details on colouring and notation, we refer to section \ref{sec:Notation}.}
    \label{fig:Dp-NS5-Configurations-2}
\end{figure}

\clearpage
\subsection{P/F1 - NS5/KKM configurations}

\begin{figure}[ht!]
    \centering
    \includegraphics[width = 0.5 \textwidth]{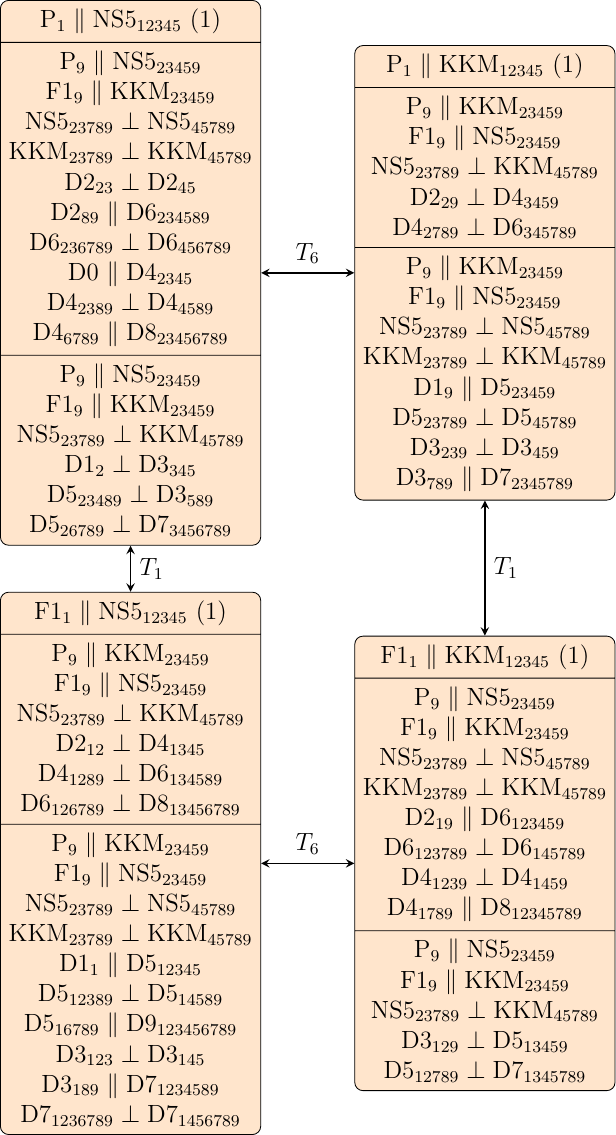}
    \caption{Possible local supersymmetry enhancements of $\W$/$\F$-$\NS$/$\KKM$ standard configurations. For details on colouring and notation, we refer to section \ref{sec:Notation}.}
    \label{fig:P-NS5-Configurations}
\end{figure}

\clearpage
\subsection{NS5/KKM - NS5/KKM configurations}


\begin{figure}[ht!]
    \centering
    \includegraphics[width = 1.0 \textwidth]{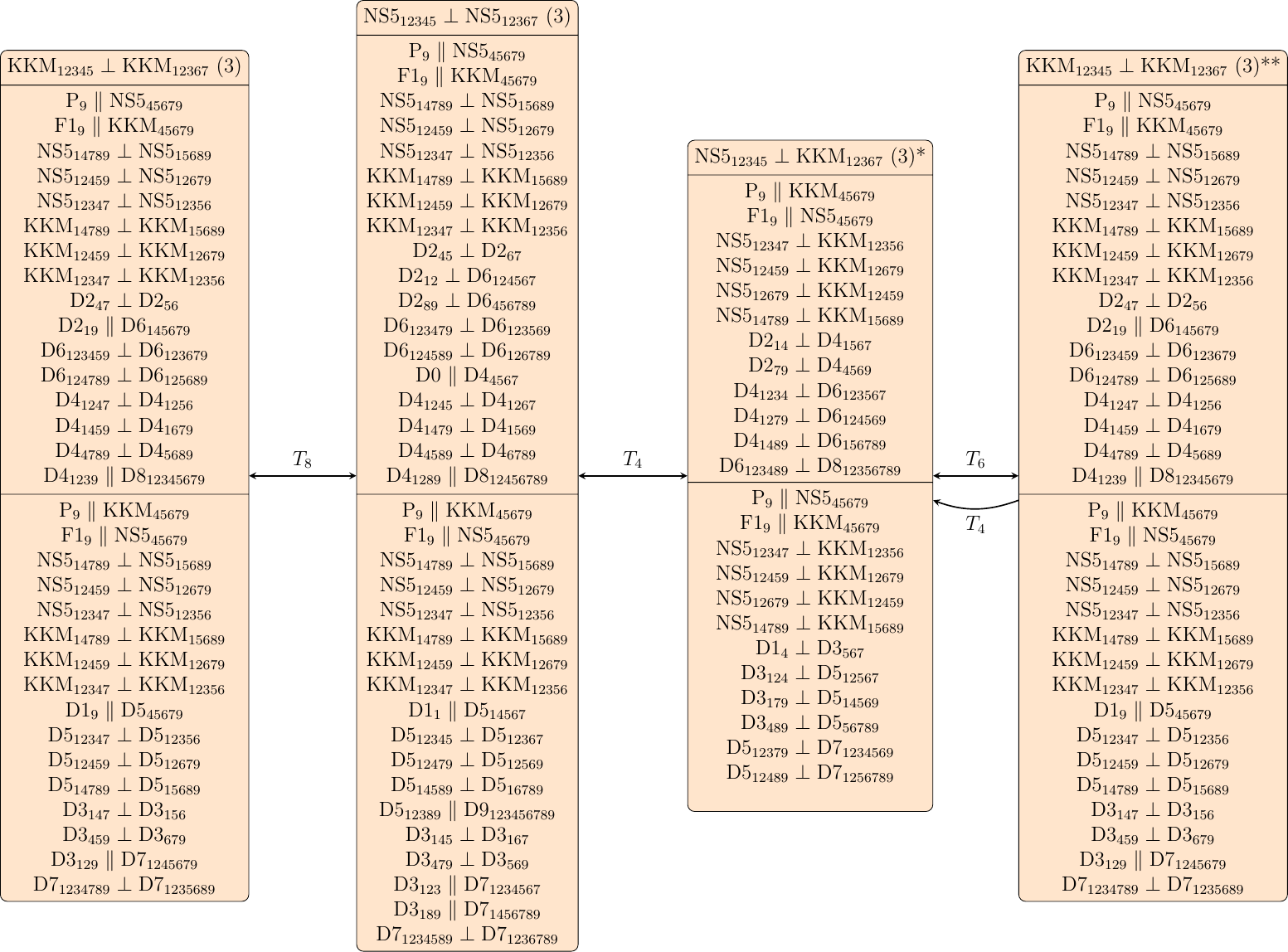}
    \caption{Possible local supersymmetry enhancements of $\NS$/$\KKM$-$\NS$/$\KKM$ standard configurations. For details on colouring and notation, we refer to section \ref{sec:Notation}.}
    \label{fig:NS5-KKM-Configurations-reordered}
\end{figure}

\clearpage
\subsection{Dp - Dq non-standard configurations}

\begin{figure}[ht!]
    \centering
    \includegraphics[width = 1.0 \textwidth]{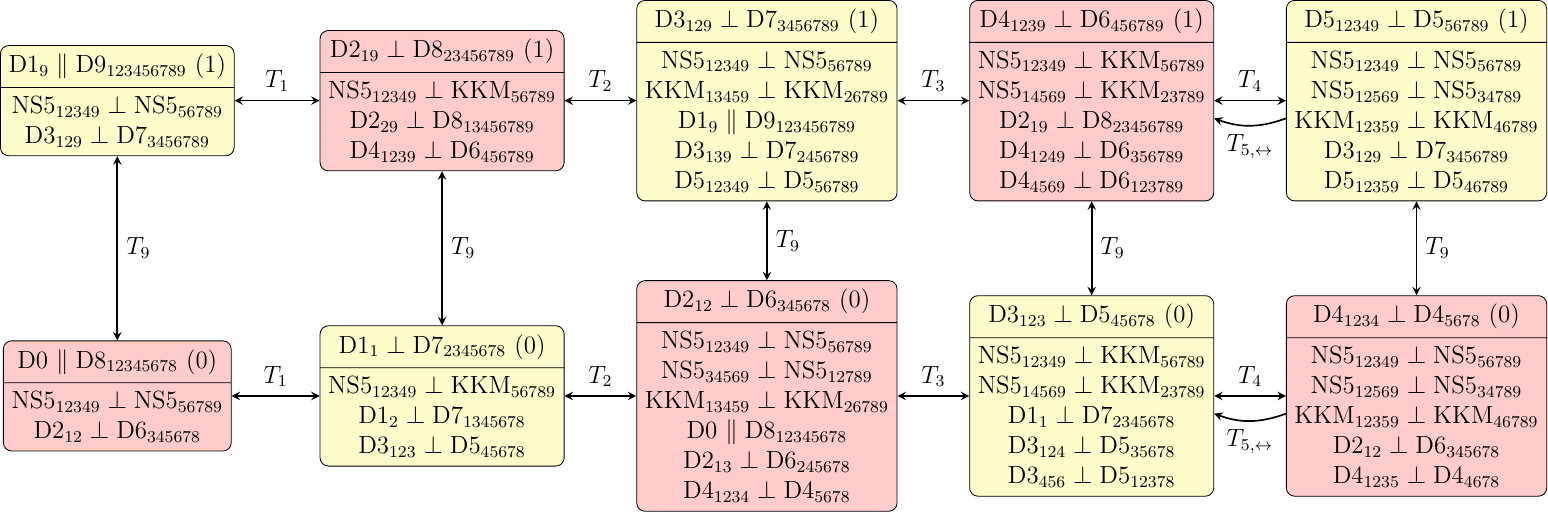}
    \caption{Possible local supersymmetry enhancements of $\D{p}$-$\D{q}$ non-standard configurations. For details on colouring and notation, we refer to section \ref{sec:Notation}.}
    \label{fig:Dp-Dq-NonStandard-Configurations}
\end{figure}

\clearpage
\subsection{NS5/KKM - NS5/KKM non-standard configurations}


\begin{figure}[ht!]
    \centering
    \includegraphics[width = 1.0 \textwidth]{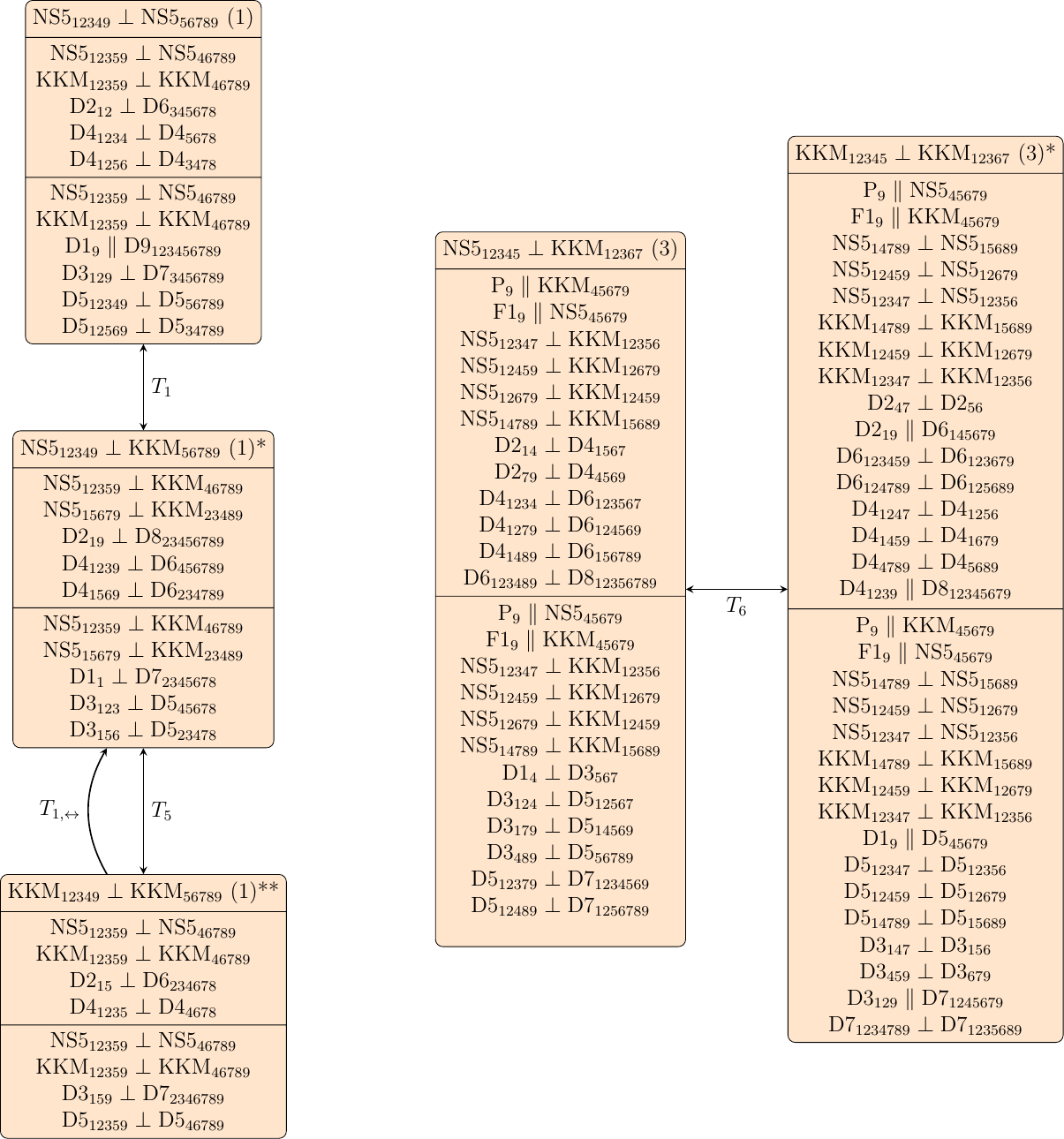}
    \caption{Possible local supersymmetry enhancements of $\NS$/$\KKM$-$\NS$/$\KKM$ non-standard configurations. For details on colouring and notation, we refer to section \ref{sec:Notation}.}
    \label{fig:NS5-KKM-NonStandard-Configurations-reordered}
\end{figure}

\clearpage
\subsection{Dp - NS5 and Dq - KKM non-standard configurations}


\begin{figure}[ht!]
    \centering
    \includegraphics[width = 0.8 \textwidth]{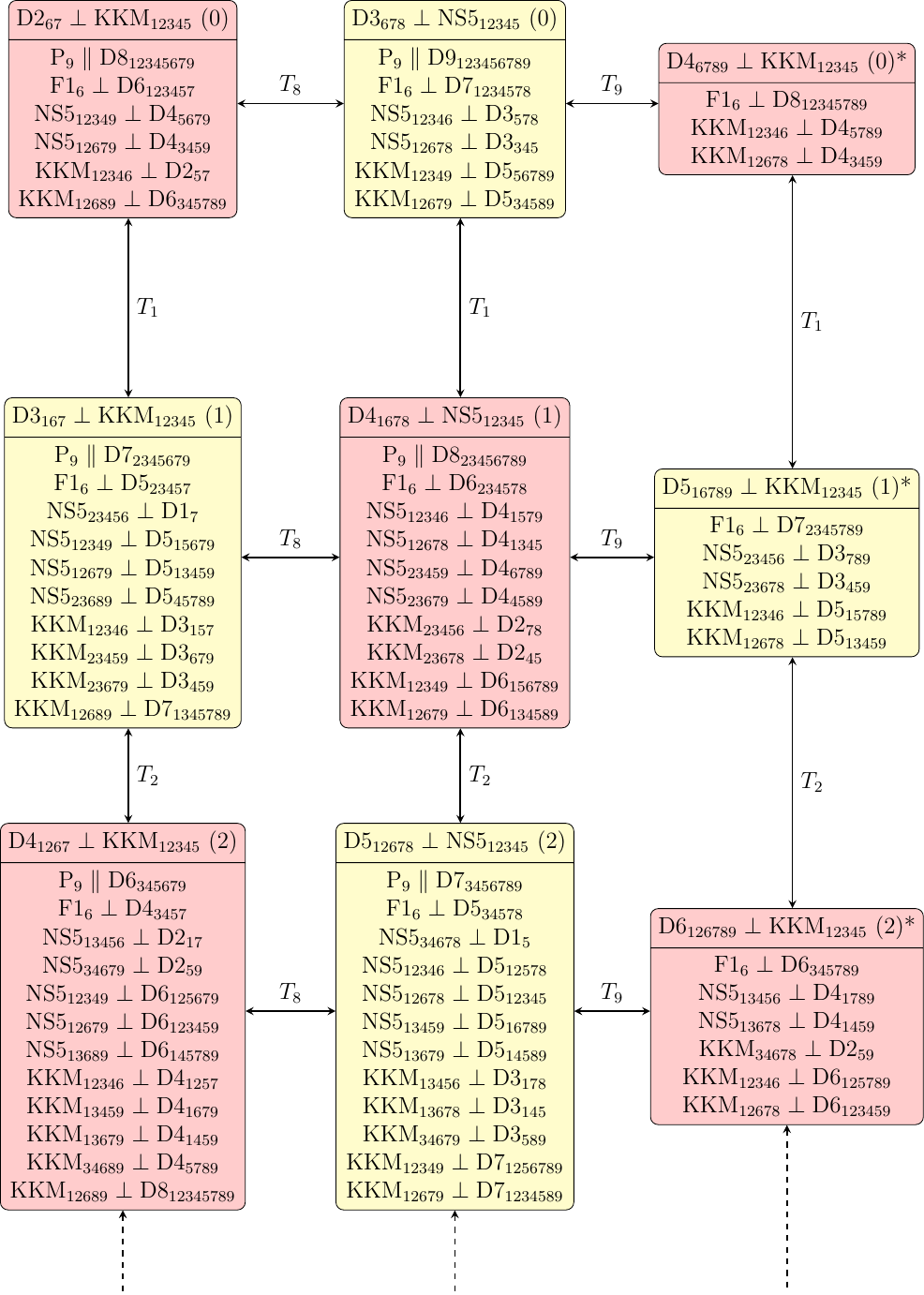}
    \caption{Possible local supersymmetry enhancements of $\D{p}$-$\NS$ and $\D{q}$-$\KKM$ non-standard configurations (first part). For details on colouring and notation, we refer to section \ref{sec:Notation}.}
    \label{fig:Dp-NS5-NonStandard-Configurations-1}
\end{figure}
\begin{figure}[ht!]
    \centering
    \includegraphics[width = 0.8 \textwidth]{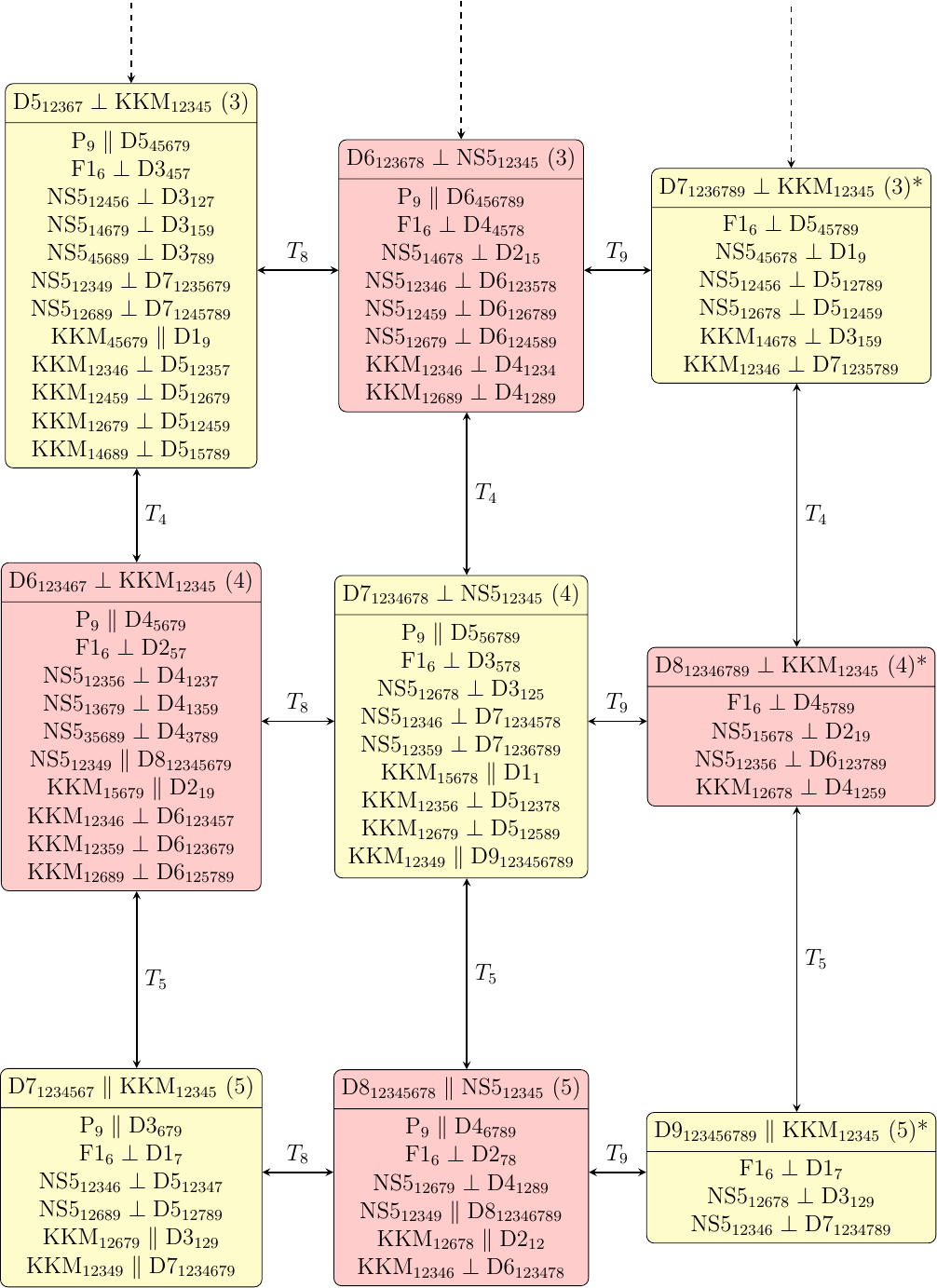}
    \caption{Possible local supersymmetry enhancements of $\D{p}$-$\NS$ and $\D{q}$-$\KKM$ non-standard configurations (second part). For details on colouring and notation, we refer to section \ref{sec:Notation}.}
    \label{fig:Dp-NS5-NonStandard-Configurations-2}
\end{figure}

\end{appendix}

  \backmatter
  \bibliographystyle{JHEP}
  \bibliography{literature}

\providecommand{\href}[2]{#2}\begingroup\raggedright\begin{thebibliography}{10}

\bibitem{Bena:2011uw}
I.~Bena, J.~de~Boer, M.~Shigemori and N.~P. Warner, \emph{{Double, Double Supertube Bubble}}, \href{https://doi.org/10.1007/JHEP10(2011)116}{\emph{JHEP} {\bfseries 10} (2011) 116} [\href{https://arxiv.org/abs/1107.2650}{{\ttfamily 1107.2650}}].

\bibitem{Hawking:1974rv}
S.~W. Hawking, \emph{{Black hole explosions}}, \href{https://doi.org/10.1038/248030a0}{\emph{Nature} {\bfseries 248} (1974) 30}.

\bibitem{Bekenstein:1972tm}
J.~D. Bekenstein, \emph{{Black holes and the second law}}, \href{https://doi.org/10.1007/BF02757029}{\emph{Lett. Nuovo Cim.} {\bfseries 4} (1972) 737}.

\bibitem{Hawking:1975vcx}
S.~W. Hawking, \emph{{Particle Creation by Black Holes}}, \href{https://doi.org/10.1007/BF02345020}{\emph{Commun. Math. Phys.} {\bfseries 43} (1975) 199}.

\bibitem{Israel:1967}
W.~Israel, \emph{Event horizons in static vacuum space-times}, \href{https://doi.org/10.1103/PhysRev.164.1776}{\emph{Phys. Rev.} {\bfseries 164} (1967) 1776}.

\bibitem{Israel:1968}
W.~Israel, \emph{Event horizons in static electrovac space-times}, \href{https://doi.org/10.1007/BF01645859}{\emph{Communications in Mathematical Physics} {\bfseries 8} (1968) 245}.

\bibitem{Carter:1971}
B.~Carter, \emph{Axisymmetric black hole has only two degrees of freedom}, \href{https://doi.org/10.1103/PhysRevLett.26.331}{\emph{Phys. Rev. Lett.} {\bfseries 26} (1971) 331}.

\bibitem{Maeda:2011sh}
K.-i. Maeda and M.~Nozawa, \emph{{Black hole solutions in string theory}}, \href{https://doi.org/10.1143/PTPS.189.310}{\emph{Prog. Theor. Phys. Suppl.} {\bfseries 189} (2011) 310} [\href{https://arxiv.org/abs/1104.1849}{{\ttfamily 1104.1849}}].

\bibitem{Tseytlin:1996bh}
A.~A. Tseytlin, \emph{{Harmonic superpositions of M-branes}}, \href{https://doi.org/10.1016/0550-3213(96)00328-8}{\emph{Nucl. Phys. B} {\bfseries 475} (1996) 149} [\href{https://arxiv.org/abs/hep-th/9604035}{{\ttfamily hep-th/9604035}}].

\bibitem{Horowitz:1996nw}
G.~T. Horowitz and J.~Polchinski, \emph{{A Correspondence principle for black holes and strings}}, \href{https://doi.org/10.1103/PhysRevD.55.6189}{\emph{Phys. Rev. D} {\bfseries 55} (1997) 6189} [\href{https://arxiv.org/abs/hep-th/9612146}{{\ttfamily hep-th/9612146}}].

\bibitem{Strominger:1996sh}
A.~Strominger and C.~Vafa, \emph{{Microscopic origin of the Bekenstein-Hawking entropy}}, \href{https://doi.org/10.1016/0370-2693(96)00345-0}{\emph{Phys. Lett. B} {\bfseries 379} (1996) 99} [\href{https://arxiv.org/abs/hep-th/9601029}{{\ttfamily hep-th/9601029}}].

\bibitem{Maldacena:1996gb}
J.~M. Maldacena and A.~Strominger, \emph{{Statistical entropy of four-dimensional extremal black holes}}, \href{https://doi.org/10.1103/PhysRevLett.77.428}{\emph{Phys. Rev. Lett.} {\bfseries 77} (1996) 428} [\href{https://arxiv.org/abs/hep-th/9603060}{{\ttfamily hep-th/9603060}}].

\bibitem{Johnson:1996ga}
C.~V. Johnson, R.~R. Khuri and R.~C. Myers, \emph{{Entropy of 4-D extremal black holes}}, \href{https://doi.org/10.1016/0370-2693(96)00383-8}{\emph{Phys. Lett. B} {\bfseries 378} (1996) 78} [\href{https://arxiv.org/abs/hep-th/9603061}{{\ttfamily hep-th/9603061}}].

\bibitem{Bena:2022wpl}
I.~Bena, S.~D. Hampton, A.~Houppe, Y.~Li and D.~Toulikas, \emph{{The (amazing) super-maze}}, \href{https://doi.org/10.1007/JHEP03(2023)237}{\emph{JHEP} {\bfseries 03} (2023) 237} [\href{https://arxiv.org/abs/2211.14326}{{\ttfamily 2211.14326}}].

\bibitem{Mateos:2001qs}
D.~Mateos and P.~K. Townsend, \emph{{Supertubes}}, \href{https://doi.org/10.1103/PhysRevLett.87.011602}{\emph{Phys. Rev. Lett.} {\bfseries 87} (2001) 011602} [\href{https://arxiv.org/abs/hep-th/0103030}{{\ttfamily hep-th/0103030}}].

\bibitem{Emparan:2001ux}
R.~Emparan, D.~Mateos and P.~K. Townsend, \emph{{Supergravity supertubes}}, \href{https://doi.org/10.1088/1126-6708/2001/07/011}{\emph{JHEP} {\bfseries 07} (2001) 011} [\href{https://arxiv.org/abs/hep-th/0106012}{{\ttfamily hep-th/0106012}}].

\bibitem{Bena:2013ora}
I.~Bena, S.~F. Ross and N.~P. Warner, \emph{{On the Oscillation of Species}}, \href{https://doi.org/10.1007/JHEP09(2014)113}{\emph{JHEP} {\bfseries 09} (2014) 113} [\href{https://arxiv.org/abs/1312.3635}{{\ttfamily 1312.3635}}].

\bibitem{Mathur:2013nja}
S.~D. Mathur and D.~Turton, \emph{{Oscillating supertubes and neutral rotating black hole microstates}}, \href{https://doi.org/10.1007/JHEP04(2014)072}{\emph{JHEP} {\bfseries 04} (2014) 072} [\href{https://arxiv.org/abs/1310.1354}{{\ttfamily 1310.1354}}].

\bibitem{Bena:2015bea}
I.~Bena, S.~Giusto, R.~Russo, M.~Shigemori and N.~P. Warner, \emph{{Habemus Superstratum! A constructive proof of the existence of superstrata}}, \href{https://doi.org/10.1007/JHEP05(2015)110}{\emph{JHEP} {\bfseries 05} (2015) 110} [\href{https://arxiv.org/abs/1503.01463}{{\ttfamily 1503.01463}}].

\bibitem{Seiberg:1997zk}
N.~Seiberg, \emph{{New theories in six-dimensions and matrix description of M theory on T**5 and T**5 / Z(2)}}, \href{https://doi.org/10.1016/S0370-2693(97)00805-8}{\emph{Phys. Lett. B} {\bfseries 408} (1997) 98} [\href{https://arxiv.org/abs/hep-th/9705221}{{\ttfamily hep-th/9705221}}].

\bibitem{Kutasov:2001uf}
D.~Kutasov, \emph{{Introduction to little string theory}}, {\emph{ICTP Lect. Notes Ser.} {\bfseries 7} (2002) 165}.

\bibitem{Dijkgraaf:1996cv}
R.~Dijkgraaf, E.~P. Verlinde and H.~L. Verlinde, \emph{{BPS spectrum of the five-brane and black hole entropy}}, \href{https://doi.org/10.1016/S0550-3213(96)00638-4}{\emph{Nucl. Phys. B} {\bfseries 486} (1997) 77} [\href{https://arxiv.org/abs/hep-th/9603126}{{\ttfamily hep-th/9603126}}].

\bibitem{Callan:1997kz}
C.~G. Callan and J.~M. Maldacena, \emph{{Brane death and dynamics from the Born-Infeld action}}, \href{https://doi.org/10.1016/S0550-3213(97)00700-1}{\emph{Nucl. Phys. B} {\bfseries 513} (1998) 198} [\href{https://arxiv.org/abs/hep-th/9708147}{{\ttfamily hep-th/9708147}}].

\bibitem{Eckardt:2023nmn}
B.~Eckardt and Y.~Li, \emph{{The 1/4-BPS building blocks of brane interactions}},  \href{https://arxiv.org/abs/2312.13269}{{\ttfamily 2312.13269}}.

\bibitem{Shigemori:2019orj}
M.~Shigemori, \emph{{Counting Superstrata}}, \href{https://doi.org/10.1007/JHEP10(2019)017}{\emph{JHEP} {\bfseries 10} (2019) 017} [\href{https://arxiv.org/abs/1907.03878}{{\ttfamily 1907.03878}}].

\bibitem{Smith:2002wn}
D.~J. Smith, \emph{{Intersecting brane solutions in string and M theory}}, \href{https://doi.org/10.1088/0264-9381/20/9/203}{\emph{Class. Quant. Grav.} {\bfseries 20} (2003) R233} [\href{https://arxiv.org/abs/hep-th/0210157}{{\ttfamily hep-th/0210157}}].

\bibitem{Bergshoeff:1997kr}
E.~Bergshoeff, R.~Kallosh, T.~Ortin and G.~Papadopoulos, \emph{{Kappa symmetry, supersymmetry and intersecting branes}}, \href{https://doi.org/10.1016/S0550-3213(97)00470-7}{\emph{Nucl. Phys. B} {\bfseries 502} (1997) 149} [\href{https://arxiv.org/abs/hep-th/9705040}{{\ttfamily hep-th/9705040}}].

\bibitem{Li:2023jxb}
Y.~Li, \emph{{Local supersymmetries and three-charge black holes}},  in \emph{{CORFU2022: 22th Hellenic School and Workshops on Elementary Particle Physics and Gravity}}, 5, 2023, \href{https://arxiv.org/abs/2305.03747}{{\ttfamily 2305.03747}}.

\bibitem{Dabholkar:1995nc}
A.~Dabholkar, J.~P. Gauntlett, J.~A. Harvey and D.~Waldram, \emph{{Strings as Solitons \& Black Holes as Strings}}, \href{https://doi.org/10.1016/0550-3213(96)00266-0}{\emph{Nucl. Phys.} {\bfseries B474} (1996) 85} [\href{https://arxiv.org/abs/hep-th/9511053}{{\ttfamily hep-th/9511053}}].

\bibitem{Hashimoto:2003pu}
K.~Hashimoto and W.~Taylor, \emph{{Strings between branes}}, \href{https://doi.org/10.1088/1126-6708/2003/10/040}{\emph{JHEP} {\bfseries 10} (2003) 040} [\href{https://arxiv.org/abs/hep-th/0307297}{{\ttfamily hep-th/0307297}}].

\bibitem{Hashimoto:1997px}
A.~Hashimoto, \emph{{The Shape of branes pulled by strings}}, \href{https://doi.org/10.1103/PhysRevD.57.6441}{\emph{Phys. Rev. D} {\bfseries 57} (1998) 6441} [\href{https://arxiv.org/abs/hep-th/9711097}{{\ttfamily hep-th/9711097}}].

\end{thebibliography}\endgroup
  \markboth{}{}

\end{document}